\documentclass[a4paper]{amsart}
\usepackage{xr-hyper}
\usepackage[breaklinks=true,
bookmarks=true,
backref=page,
hyperindex,
pagebackref=true,
]{hyperref}
\usepackage{graphicx}
\usepackage[backrefs,msc-links,initials,nobysame,short-journals]{amsrefs}

\BibSpecAlias{incollection}{inproceedings}
\BibSpec{article}{%
    +{}  {\PrintAuthors}                {author}
    +{.} { }                            {title}
    +{.} { }                            {part}
    +{:} { }                            {subtitle}
    +{.} { \PrintContributions}         {contribution}
    +{.} { \PrintPartials}              {partial}
    +{.} { \emph}                       {journal}
    +{,} { \textbf}                            {volume}
    +{}  { \parenthesize}               {number}
    +{:} {}                             {pages}
    +{,} { \PrintDateB}                 {date}
    +{,} { }                            {status}
    +{.} { \PrintTranslation}           {translation}
    +{.} { Reprinted in \PrintReprint}  {reprint}
    +{.} { }                            {note}
    +{.} {}                             {transition}
}

\BibSpec{partial}{%
    +{}  {}                             {part}
    +{:} { }                            {subtitle}
    +{.} { \PrintContributions}         {contribution}
    +{.} { \emph}                       {journal}
    +{,} { \textbf}                            {volume}
    +{}  { \parenthesize}               {number}
    +{:} {}                             {pages}
    +{,} { \PrintDateB}                 {date}
}

\BibSpec{book}{%
    +{}  {\PrintPrimary}                {transition}
    +{.} { \emph}                       {title}
    +{.} { }                            {part}
    +{:} { \emph}                       {subtitle}
    +{.} { }                            {series}
    +{,} { \voltext}                    {volume}
    +{.} { Edited by \PrintNameList}    {editor}
    +{.} { Translated by \PrintNameList}{translator}
    +{.} { \PrintContributions}         {contribution}
    +{.} { }                            {publisher}
    +{.} { }                            {organization}
    +{,} { }                            {address}
    +{,} { \PrintEdition}               {edition}
    +{,} { \PrintDateB}                 {date}
    +{.} { }                            {note}
    +{.} {}                             {transition}
    +{.} { \PrintTranslation}           {translation}
    +{.} { Reprinted in \PrintReprint}  {reprint}
    +{.} {}                             {transition}
}

\BibSpec{collection.article}{%
    +{}  {\PrintAuthors}                {author}
    +{.} { }                            {title}
    +{.} { }                            {part}
    +{:} { }                            {subtitle}
    +{.} { \PrintContributions}         {contribution}
    +{.} { \PrintConference}            {conference}
    +{.} { \PrintBook}                  {book}
    +{.} { In \PrintEditorsA}              {editor}
    +{.} { \emph}                         {booktitle}
    +{,} { pages~}                      {pages}
    +{,} { }                            {publisher}
    +{,} { }                            {organization}
    +{,} { }                            {address}
    +{,} { \PrintDateB}                 {date}
    +{.} { \PrintTranslation}           {translation}
    +{.} { Reprinted in \PrintReprint}  {reprint}
    +{.} { }                            {note}
    +{.} {}                             {transition}
}

 \BibSpec{inproceedings}{%
     +{}  {\PrintAuthors}                {author}
     +{.} { }                            {title}
     +{.} { }                            {part}
     +{:} { }                            {subtitle}
     +{.} { \PrintContributions}         {contribution}
     +{.} { \PrintConference}            {conference}
     +{.} { \PrintBook}                  {book}
     +{.} { In \PrintEditorsA}              {editor}
     +{.} {  \emph}                         {booktitle}
     +{,} { pages~}                      {pages}
     +{,} { }                            {publisher}
     +{,} { }                            {organization}
     +{,} { }                            {address}
     +{,} { \PrintDateB}                 {date}
     +{.} { \PrintTranslation}           {translation}
     +{.} { Reprinted in \PrintReprint}  {reprint}
     +{.} { }                            {note}
     +{.} {}                             {transition}
 }

\BibSpec{conference}{%
    +{}  {}                        {title}
    +{}  {\PrintConferenceDetails} {transition}
}

\BibSpec{innerbook}{%
    +{.} { \emph}                       {title}
    +{.} { }                            {part}
    +{:} { \emph}                       {subtitle}
    +{.} { }                            {series}
    +{,} { \voltext}                    {volume}
    +{.} { Edited by \PrintNameList}    {editor}
    +{.} { Translated by \PrintNameList}{translator}
    +{.} { \PrintContributions}         {contribution}
    +{.} { }                            {publisher}
    +{.} { }                            {organization}
    +{,} { }                            {address}
    +{,} { \PrintEdition}               {edition}
    +{,} { \PrintDateB}                 {date}
    +{.} { }                            {note}
    +{.} {}                             {transition}
}

\BibSpec{report}{%
    +{}  {\PrintPrimary}                {transition}
    +{.} { \emph}                       {title}
    +{.} { }                            {part}
    +{:} { \emph}                       {subtitle}
    +{.} { \PrintContributions}         {contribution}
    +{.} { Technical Report }           {number}
    +{,} { }                            {series}
    +{.} { }                            {organization}
    +{,} { }                            {address}
    +{,} { \PrintDateB}                 {date}
    +{.} { \PrintTranslation}           {translation}
    +{.} { Reprinted in \PrintReprint}  {reprint}
    +{.} { }                            {note}
    +{.} {}                             {transition}
}

\BibSpec{thesis}{%
    +{}  {\PrintAuthors}                {author}
    +{,} { \emph}                       {title}
    +{:} { \emph}                       {subtitle}
    +{.} { \PrintThesisType}            {type}
    +{.} { }                            {organization}
    +{,} { }                            {address}
    +{,} { \PrintDateB}                 {date}
    +{.} { \PrintTranslation}           {translation}
    +{.} { Reprinted in \PrintReprint}  {reprint}
    +{.} { }                            {note}
    +{.} {}                             {transition}
}
\usepackage{amssymb}
\usepackage{color}
\newtheorem{thm}{Theorem}[section]
   \PassOptionsToPackage{british}{babel}
    \newtheorem{prop}[thm]{Proposition}
    
    \newtheorem{cor}[thm]{Corollary}
   \theoremstyle{definition}

    \newtheorem{defn}[thm]{Definition}

    \newtheorem{example}[thm]{Example}

   \theoremstyle{remark}
    \newtheorem{rem}[thm]{Remark}
    
\providecommand{\norm}[2][\relax]{\left\|#2\right\|\ifx#1\relax\else_{#1}\fi}
\providecommand{\modulus}[2][\relax]{\left| #2 \right|\ifx#1\relax\else_{#1}\fi}
\providecommand{\oper}[1]{\mathcal{#1}}
\providecommand{\algebra}[1]{\ensuremath{\mathfrak{#1}}}

\providecommand{\Space}[3][]{\ifx#2R\ifx#1e \mathbb{C}^{#3} \else
\ifx#1p \mathbb{D}^{#3} \else
\ifx#1h \mathbb{O}^{#3} \else
\ifx#1\sigma \mathbb{A}\!^{#3} \else
\ensuremath{\mathbb{#2}^{#3}_{#1}{}} \fi \fi \fi \fi \else
\ensuremath{\mathbb{#2}^{#3}_{#1}{}} \fi}
\providecommand{\FSpace}[3][]{\ensuremath{\ifx#2l \ell_{#3}^{#1}{}\else
  \mathsf{#2}_{#3}^{#1}{}\fi}}

\providecommand{\uir}[3][0]{\ifcase #1{\rho^{#2}_{#3}}%
\or {\breve{\rho}^{#2}_{#3}}%
\or {\tilde{\rho}^{#2}_{#3}}\fi}
\providecommand{\scalar}[3][\relax]{\left\langle #2,#3
        \right\rangle\ifx#1\relax\else_{#1}\fi}

\providecommand{\SL}[1][2]{\ensuremath{\mathrm{SL}_{#1}(\Space{R}{})}}

\providecommand{\rmi}{\mathrm{i}}
\providecommand{\rme}{\mathrm{e}}
\providecommand{\rmd}{\mathrm{d}}

\providecommand{\linv}[2][\relax]{\mathfrak{L}^{#2}_{#1}}
\providecommand{\myhbar}{\hslash}
\providecommand{\myh}{h}
\providecommand{\map}[1]{\mathsf{#1}}

\providecommand{\Zbl}[1]{Zbl\href{http://www.emis.de:80/cgi-bin/zmen/ZMATH/en/zmathf.html?first=1&maxdocs=3&type=html&an=#1&format=complete}{#1}}

\providecommand{\myeprint}[2]{\href{#1}{\texttt{#2}}}

\numberwithin{equation}{section}

\begin{document}
\title[Harmonic Oscillator and Squeezed States]{Geometric Dynamics of
  a Harmonic Oscillator,\\  Arbitrary Minimal Uncertainty States\\
  and  the Smallest Step \(3\) Nilpotent Lie Group}

\author[F. Almalki and V.V. Kisil]
{Fadhel Almalki and \href{http://www.maths.leeds.ac.uk/~kisilv/}{Vladimir V. Kisil}}
\thanks{On  leave from Odessa University.}

\address{%
School of Mathematics\\
University of Leeds\\
Leeds LS2\,9JT\\
UK
}

\email{\href{mailto:kisilv@maths.leeds.ac.uk}{kisilv@maths.leeds.ac.uk}}

\urladdr{\url{http://www.maths.leeds.ac.uk/~kisilv/}}

\date{\today}

\begin{abstract}
  The paper presents a new method of geometric solution of a
  Schr\"odinger equation by constructing an equivalent first-order
  partial differential equation with a bigger number of variables. The
  equivalent equation shall be restricted to a specific subspace with
  auxiliary conditions which are obtained from a coherent state
  transform.  The method is applied to the fundamental case of the
  harmonic oscillator and coherent state transform generated by the
  minimal nilpotent step three Lie group---the group \(\mathbb{G}\)
  (also known under many names, e.g. quartic group). We obtain a
  geometric solution for an arbitrary minimal uncertainty state used
  as a fiducial vector. In contrast, it is shown that the well-known
  Fock--Segal--Bargmann transform and the Heisenberg group require a
  specific fiducial vector to produce a geometric solution. A
  technical aspect considered in this paper is that a certain
  modification of a coherent state transform is required: although the
  irreducible representation of the group \(\mathbb{G}\) is
  square-integrable modulo a subgroup \(H\), the obtained dynamic is
  transverse to the homogeneous space \(\mathbb{G}/H\).
\end{abstract}
\keywords{}
\subjclass[2010]{Primary 81R30; Secondary 20C35, 22E70, 35Q70, 35A25, 81V80.}
\dedicatory{Dedicated to 70-th birthday of Maurice de Gosson}
\maketitle
\tableofcontents

\section{Introduction}
\label{sec:introduction}

A transition from configuration space to the phase space in quantum
mechanics is performed by the coherent states
transform~\cites{deGosson11a, deGosson06a, Folland89}. As a result the
number of independent variables is doubled from \(\Space{R}{n}\) to
\(\Space{C}{n}\) and an irreducible Fock--Segal--Bargmann space is
characterised by the analyticity
condition~\cites{Vourdas06a,Kisil17a}. The dynamic of quantum harmonic
oscillator in Fock--Segal--Bargmann space is given geometrically as
one-parameter subgroup of symplectomorphisms. In this paper we
describe a method generalising this construction of geometric
solutions of the Schr\"odinger equation.

The technique is applied to obtain a novel picture of
a harmonic oscillator dynamics associated to squeezed states.
The harmonic oscillator is an archetypal example of exactly
solvable classical and quantum systems with applications to many
branches of physics, e.g. optics~\cite{Wolf10a}.
However, not only the solution of harmonic oscillator has numerous
uses, important methods with a wide applicability were initially
tested on this fundamental case. The most efficient solution of the
spectral problem through \emph{ladder operators} were successfully
expanded from the harmonic oscillator to hydrogen
atom~\cite{Schrodinger40a} with further generalisations in
supersymmetric quantum mechanics~\cite{CarinenaPlyushchay17a} and
pseudo-boson framework~\cite{AliBagarelloGazeau15a}.

In order to clarify our target we need to recall briefly the scheme
with an emphasis on some details which are usually depressed in
abridged presentations. Let the quantum observables for coordinate
\(q\) and momentum \(p\) satisfy the Heisenberg canonical commutator
relation (CCR):
\begin{equation}
  \label{eq:heisenberg-commutator-CCR}
  [q,p]=\rmi\myhbar I\,.
\end{equation}
Then, for a harmonic oscillator \emph{with the mass \(m\) and frequency
  \(\omega\)} one introduces the ladder (annihilation and creation) operators \((a^{-},a^{+})\):
\begin{equation}
  \label{eq:ladder-heisenberg}
  a^{\pm}=
  \sqrt{\frac{m\omega}{2\myhbar}}\left(q\mp \frac{\rmi}{m\omega}
    p\right) \quad\text{ with the commutator } 
  [a^-,a^+]=I.
\end{equation}
Thereafter, the Hamiltonian \(H\) of the harmonic oscillator can be
expressed through the ladder operators:
\begin{equation}
  \label{eq:harmonic-hamiltonian}
  H=\frac{1}{2}\left(\frac{1}{m}p^2+m \omega^2 q^2\right)= \myhbar\omega \left(a^+a^-+\frac{1}{2}\right)\,.
\end{equation}
Assuming the existence of a non-zero vacuum \(\phi_0\) such that
\(a^{-}\phi_0 =0\), the commutator
relations~\eqref{eq:ladder-heisenberg} imply that states
\(\phi_n=(a^{+})^n \phi_0\) are eigenvectors of \(H\) with eigenvalues
\(n+\frac{1}{2}\). This completely solves the spectral problem for the
harmonic oscillator, but a couple of observations are in place:
\begin{enumerate}
\item The ladder operators~\eqref{eq:ladder-heisenberg} (and
  subsequently the eigenvectors \(\phi_n\)) depend on the parameter
  \(m\omega\). They \emph{are not useful} for a harmonic oscillator with
  a different value of \(m'\omega'\).
\item The explicit dynamic of an arbitrary state \(f\) \emph{is not
    transparent}: first we need to find its decomposition
  \(f=\sum_n c_n \phi_n\) over the orthogonal basis of eigenvectors
  \(\phi_n\) and then express evolution as
  \begin{equation}
    \label{eq:dynamics-in-eigenvector}
    F(t)=\sum_n c_n \rme^{-\rmi (n+\frac{1}{2}) \omega t} \phi_n\,.
  \end{equation}
\end{enumerate}
The latter can be fixed by the coherent states transform to
Fock--Segal--Bargmann (FSB) space~\cite{Folland89}*{1.6}:
\begin{equation}
  \label{eq:wavelet-transform}
  \oper{W}_{\phi_0}:\; f \rightarrow \tilde{f}(z)=\scalar{f}{\phi_z}\,,\qquad
  \text{where }\phi_z=\rme^{za^+ -\bar{z}a^-}\phi_0
  \quad \text{and}\quad z\in\Space{C}{}\,.
\end{equation}
This presents the dynamic of a state \(\tilde{f}(z)\) for the
Hamiltonian \(H\)~\eqref{eq:harmonic-hamiltonian} in a geometric
fashion, cf.~\eqref{eq:dynamics-in-eigenvector}:
\begin{equation}
  \label{eq:harmonic-oscillator-FSB}
  F(t;z)=\rme^{-\pi \rmi \omega t} \tilde{f}\left(\rme^{-2\pi \rmi\omega t}z\right), \qquad \text{where } \tilde{f}(z)=F(0;z)\,.
\end{equation}
Yet, this presentation still relays on the vacuum \(\phi_0\) (and,
thus, all other coherent states
\(\phi_z\)~\eqref{eq:wavelet-transform}) having the given value of
\(m\omega\) as before. Metaphorically, the traditional usage of the
ladder operators and vacuum \(\phi_0\) is like a key, which can unlock
only the matching harmonic oscillator with the same value of
\(m\omega\).

On the other hand, the collection of all vacuums \(\phi_0\) for
various values of \(m\omega\) form a distinguished class of
\emph{minimal uncertainty states} (aka \emph{squeezed
  states})~\citelist{\cite{Walls08}*{\S\,2.4--2.5}
  \cite{Schleich01a}*{\S\,4.3} \cite{Gazeau09a}*{Ch.~10}
  \cite{deGosson11a}*{\S\,11.3}}, which have the smallest allowed
value:
\begin{displaymath}
  \Delta_{\phi_0}(p) \cdot \Delta_{\phi_0}(q) =\frac{\myhbar}{2}\,.
\end{displaymath}
In this paper we present \emph{an extension of the traditional
  framework, which allows to use any minimal uncertainty state
  \(\phi_0\) for a harmonic oscillator with a different value of
  \(m\omega\)}:
\begin{itemize}
\item to obtain a geometric dynamic similar
  to~\eqref{eq:harmonic-oscillator-FSB}, cf.
  \S\,\ref{sec:harm-oscill-from-shear};
\item to create eigenvectors, cf. Sect.~\ref{hamiltonian and ladder}.
\end{itemize}

Our construction is based on the extension of the Heisenberg group
\(\Space{H}{}\)~\citelist{ \cite{Folland89}*{Ch.~1}
  \cite{Kirillov04a}*{Ch.~2} \cite{Kisil11c} \cite{Schempp86a}} to the
minimal nilpotent step \(3\) group \(\Space{G}{}\) presented in
\S\,\ref{sec:heis-shear-group}.
\begin{rem}[Background and names]
  \label{re:background-names}
  Being the simplest nilpotent step \(3\) Lie group, \(\Space{G}{}\)
  is a natural test ground for various constructions in representation
  theory~\citelist{\cite{Kirillov04a}*{\S\,3.3}
    \cite{CorwinGreenleaf90a}*{Ex.~1.3.10}} and harmonic
  analysis~\cites{HoweRatcliffWildberger84, BeltitaBeltitaPascu13a}.
  The group \(\Space{G}{}\) was called quartic group
  in~\cites{AllenAnastassiouKlink97,JorgensenKlink85,Klink94} due to
  relation to quartic anharmonic oscillator, yet the same
  paper~\cite{Klink94} shows that \(\Space{G}{}\) is also linked to
  the heat equation and charged particles in curved magnetic
  field. Some authors, e.g.~\cite{ArdentovSachkov17a}, call
  \(\Space{G}{}\) the Engel group, however an Engel group can also
  mean any group such that each element has the ``Engel property''. As
  we were pointed by an anonymous referee, the group \(\Space{G}{}\)
  was called the para-Galilei group in~\cite{BacryLevy-Leblond68a}
  (cf. \S\,\ref{sec:shear-group-schr}), but the Galilei interpretation
  is not very relevant for our particular construction. The
  Heisenberg--Weyl algebra and the Lie algebra
  \(\algebra{g}\)~\eqref{anh comutation relation} are the first two
  terms of the infinite series of {filiform
    algebras}~\cite{Vergne70a}. A fully descriptive name of the group
  \(\Space{G}{}\) can be ``the Heisenberg group extended by shear
  transformations'' (cf. \S\,\ref{sec:shear-group-schr}), but this may
  lead to a confusion with unrelated
  shearlets~\cite{KutyniokLabate12a}. Since we are not satisfied by
  any existing name and do not want to contribute to a mix-up coining
  our own one, we will use ``the group \(\Space{G}{}\)'' throughout
  this paper.
\end{rem}
Relevantly, the group \(\Space{G}{}\) is a subgroup of the semi-direct
product \(\Space{S}{}=\Space{H}{}\rtimes_A \SL\)---the three-dimensional
Heisenberg group and the special linear group, where \(A\) is a
symplectic automorphism of \(\Space{H}{}\), see
\S\,\ref{sec:shear-group-schr}.  Such a semi-direct product arises,
for instance, in connection with the symmetry algebra based approach
of the solution of a class of parabolic differential equations: the
heat equation and the Schr\"odinger equation
\cites{Niederer72a,KalninsMiller74a,ATorre09a,Wolf76a}. In particular,
\(\Space{S}{}\) is the group of symmetries of the harmonic
oscillator~\cites{Niederer73a,Wolf76a,AldayaGuerrero01a}
and any (even time-dependent) quadratic
potential~\cite{AldayaCossioGuerreroLopez-Ruiz11b}.

A significant difference between \(\Space{H}{}\) and \(\Space{G}{}\)
is that the representations of \(\Space{G}{}\) (cf.
\S\,\ref{sec:induc-repr-homog}) are not square-integrable modulo its
centre. Although the representation of \(\Space{G}{}\) is
square-integrable modulo some subgroup \(H\), the obtained geometric
dynamic is not confined within the homogeneous space
\(\Space{G}{}/H\), cf.~\eqref{solution to the shrodinger equation for
  shear case}. Thus, the standard approach to coherent state transform
from square-integrable group
representations~\cite{AliAntGaz14a}*{Ch.~8} shall be replaced by
treatments of non-admissible mother
wavelets~\cites{Kisil09d,Kisil10c,Kisil98a}, see
also~\cites{FeichGroech89a,FeichGroech89b,Zimmermann06,Guerrero18}. This
is discussed in \S\,\ref{sec:induc-wavel-transf-shear}. We also remark
that the coherent state transform constructed in this section is closely
related to Fourier--Fresnel transform derived from the Heisenberg
group representations in~\cite{Osipov92a}.

The Fock--Segal--Bargmann space consists of analytic
functions~\citelist{ \cite{Vourdas06a} \cite{Kisil17a}
  \cite{Folland89}*{\S\,1.6}}. We revise the representation theory
background of this property in \S\,\ref{sec:char-image-space} and
deduce corresponding description of the image space of coherent state
transform on \(\Space{G}{}\). Besides the analyticity-type condition,
which relays on a suitable choice of the fiducial vector, we find an
additional condition, referred to as \emph{structural condition}, which
is completely determined by the Casimir operator of \(\Space{G}{}\) and
holds for any coherent state transform. Intriguingly, the structural
condition coincides with the Schr\"odinger equation of a free
particle. Thereafter, the image space of the coherent state transform
is obtained from FSB space through a solution of an initial value
problem.

The new method is used to analyse the harmonic oscillator in
Sect.~\ref{sec:harm-oscill-thro}.  To warm up we consider the case of
the Heisenberg group first in \S\,\ref{sec:harm-oscill-heisenberg} and
confirm that the geometric dynamic~\eqref{eq:harmonic-oscillator-FSB} is
the only possibility for the fiducial vector \(\phi_0\) with the matching
value of \(m\omega\). Then, \S\,\ref{sec:harm-oscill-from-shear}
reveals the gain from the larger group \(\Space{G}{}\): any minimal
uncertainty state can be used as a fiducial vector for a
geometrisation of dynamic. This produces an abundance of
non-equivalent coherent state transforms each delivering a time
evolution in terms of coordinate transformations~\eqref{solution to
  the shrodinger equation for shear case}. It turns out that there are
natural physical bounds of how much squeeze can be applied for a
particular state, see \S\,\ref{sec:geom-phys-mean}.

Finally we return to creation and annihilation operators in
Sect.~\ref{hamiltonian and ladder}. Predictably, their action in terms
of the group \(\Space{G}{}\) is still connected to Hermite polynomials
familiar from the standard treatment eigenfunctions of the harmonic
oscillator. This can be compared with ladder operators related to
squeezed states in~\cite{AliGorskaHorzelaSzafraniec14}.

\begin{rem}
  Interestingly, our method of order reduction for partial
  differential equations  is conceptually similar to the method  of
  order reduction of algebraic equations in Lie spheres
  geometry~\cites{FillmoreSpringer00a,Kisil14b}.
\end{rem}

\section{Preliminaries on group representations}
\label{sec:prel-group-repr}

We briefly provide main results (with further references) required for
our consideration. 

\subsection{The Heisenberg and the shear group $\Space{G}{}$}
\label{sec:heis-shear-group}

The Stone-von Neumann theorem \citelist{ \cite{Folland89}*{\S\,1.5}
  \cite{Kirillov04a}*{\S\,2.2.6}} ensures that
CCR~\eqref{eq:heisenberg-commutator-CCR} provides a representation of
the Heisenberg--Weyl algebra---the nilpotent step \(2\) Lie algebra of
the Heisenberg group \(\Space{H}{}\)~\citelist{
  \cite{Folland89}*{Ch.~1} \cite{Kirillov04a}*{Ch.~2} \cite{Kisil11c}
  \cite{Schempp86a}}. In the \emph{polarised coordinates} \((x,y,s)\)
on \(\Space{H}{}\sim \Space{R}{3}\) the group law
is~\cite{Folland89}*{\S\,1.2}:
\begin{align}
  \label{group law of the Heisenberg group}
  \textstyle (x,y,s)(x',y',s')=( x+x',y+y',s+s'+x y').
\end{align}

Let \(\algebra{g}\) be the nilpotent step \(3\) Lie algebra whose
basic elements are \(\{X_1,X_2,X_3,X_4\}\) with the following
non-vanishing
commutators~\citelist{\cite{CorwinGreenleaf90a}*{Ex.~1.3.10}
  \cite{Kirillov04a}*{\S\,3.3}}:
\begin{equation}
  \label{anh comutation relation}
  [X_1,X_2]=X_3,\qquad [X_1,X_3]=X_4\,.
\end{equation}
Clearly, the basic element corresponding to the centre of such a Lie
algebra is \(X_{4}\). Elements \(X_1\), \(X_3\) and \(X_4\) are
spanning the above mentioned Heisenberg--Weyl algebra.

The corresponding Lie group \(\Space{G}{}\) is  nilpotent step \(3\)
and its group law is:
\begin{align}
\label{group law of shear group}
  (x_1,x_2,x_3,x_4) (y_1,y_2,y_3,y_4)&=
  (x_1+y_1,x_2+y_2,x_3+y_3+ x_1 y_2,\\
  \nonumber
   &\qquad  \textstyle  x_4+y_4+x_1y_3+\frac{1}{2}  x_1^{2} y_2),
\end{align}
where \(x_j, y_j \in \Space{R}{}\) and
\((x_1,x_2,x_3,x_4):=\exp(x_4X_4)\exp(x_3X_3)\exp(x_2X_2)\exp(x_1X_1)\),
known as \emph{canonical coordinates}
~\cite{Kirillov04a}*{\S\,3.3}.

  A comparison of group laws~\eqref{group law of the
  Heisenberg group} and~\eqref{group law of shear group} shows that
the Heisenberg group \(\Space{H}{}\) is isomorphic to the subgroup
\begin{equation}
  \label{eq:heisenberg-shear}
  \tilde{ \Space{H}{}}=\{(x_1,0,x_3,x_4)\in \Space{G}{}:\  x_j \in
  \Space{R}{}\}\ \text{ by } \ (x,y,s)\mapsto
(x,0,y,s), \ (x,y,s)\in\Space{H}{}\,.
\end{equation}
In most cases we will identify \(\Space{H}{}\) and
\(\tilde{\Space{H}{}}\) through the above map. All formulae for the
Heisenberg group needed in this paper will be obtained from this
identification.

\subsection{Unitary representations of the group $\Space{G}{}$}
\label{sec:induc-repr-homog}
Unitary representations of the group \(\Space{G}{}\) can be constructed using inducing procedure (in the sense of Mackey) and Kirillov orbit method, a detailed consideration of this topic is worked out in ~\cite{Kirillov04a}*{\S\,3.3.2}
  and for an exposition of inducing procedure one may also refer to \cites{Folland95,
 KaniuthTaylor13a, Kirillov76, Kirillov04a, Varadarajan99a} with strong connections to physics~\cites{Mackey70a, Mackey85a, Mensky76,
Berndt07a} and further research
potential~\cites{Kisil17a,Kisil10a,Kisil09e}.

\begin{enumerate}
\item For 
 the centre \(Z(\Space{G}{})=\{(0,0,0,x_4)\in
  \Space{G}{}:\  x_4\in \Space{R}{}\}\) of \(\Space{G}{}\),  we have

  the following unitary representation of \(\Space{G}{}\) in
  \(\FSpace{L}{2}(\Space{R}{3})\) induced from the character \(\chi(0,0,0,x_4)=\rme^{ 2\pi\rmi \myhbar_4
    x_4}\) of the centre: 
  \begin{align}
    \label{reducible re of anhar def}
    [\tilde{\uir{}{}}(x_1,x_2,x_3,x_4)f](x_1',x_2',x_3')&=\rme^{2\pi \rmi \myhbar_4(x_4-x_1x_3+\frac{1}{2}x_1^2x_2+x_1x_3'-\frac{1}{2}x_1^2 x_2')}\\
    \nonumber
&\quad \times f(x_1'-x_1,x_2'-x_2,x_3'-x_3-x_1x_2'+x_1x_2).
  \end{align}
  This representation is reducible and we will discuss its irreducible
  components below. A restriction of~\eqref{reducible re of anhar def}
  to the Heisenberg group \(\tilde{ \Space{H}{}}\) is a variation of
  the Fock--Segal--Bargmann
  representation~\citelist{\cite{Folland89}*{\S\,1.6} \cite{Kisil17a}}. 
\item For the maximal abelian subgroup \(H_1=\{(0,x_2,x_3,x_4)\in  \Space{G}{}: x_2,x_3,x_4\in \Space{R}{}\}\), 
a generic character is parametrised by a
triple of real constants \((\myh_2,\myh_3,\myhbar_4)\) where \(\myhbar_4\)
can be identified with reduced Planck constant:
\begin{displaymath}
\chi(0,x_2,x_3,x_4)=\rme^{
2\pi\rmi(\myh_2x_2+\myh_3x_3+\myhbar_4x_4)}\\.
\end{displaymath}
The Kirillov orbit method
shows~\citelist{\cite{CorwinGreenleaf90a}*{Ex~3.1.12} \cite{Kirillov04a}*{\S\,3.3.2}}  that all
non-equivalent unitary irreducible representations are induced by
characters indexed by \((\myh_2,0,\myhbar_4)\). For such a
character the unitary representation of \(\Space{G}{}\) in
\(\FSpace{L}{2}(\Space{R}{})\) is, cf.~\cite{Kirillov04a}*{\S\,3.3, (19)}:
\begin{equation}
\label{irreducible repre of A}
[\uir{}{\myh_2 \myhbar_4}(x_1,x_2,x_3,x_4)f](x'_1)
=\rme^{ 2\pi\rmi(\myh_2x_2+   \myhbar_4(x_4-x_3x_1'+\frac{1}{2}x_2x_1'^2))}f(x'_1-x_1).
\end{equation}
This representation is indeed irreducible since its restriction to the
Heisenberg group \(\tilde{ \Space{H}{}}\) coincides with the
irreducible Schr\"odinger representation~\citelist{\cite{Folland89}*{\S\,1.3} \cite{Kisil17a}}.
\end{enumerate}

 The derived representations of \(\uir[2]{}{\myhbar_4}\), which will be
 used below, are
  \begin{align}
  \label{derived re reducibl re anharmo group}
    \rmd\uir[2]{X_1}{\myhbar_4} &= -\partial_{1}- x_2 \partial_{3} +{
      2\pi\rmi} \myhbar_4   x_3 I;&
    \rmd\uir[2]{X_2}{\myhbar_4}&= - \partial_{2};\\
    \nonumber
    \rmd\uir[2]{X_3}{\myhbar_4}&= - \partial_{3};&
    \rmd\uir[2]{X_4}{\myhbar_4}&= { 2\pi\rmi} \myhbar_4I\, .
  \end{align}
For the unitary irreducible representation \(\uir{}{\myh_2\myhbar_4}\) of \(\Space{G}{}\) in
\(\FSpace{L}{2}(\Space{R}{})\)~\eqref{irreducible repre of A} we have:
\begin{align}
\label{deri rep of irre shear group}
  \rmd\uir{X_1}{\myh_2\myhbar_4} &=-\frac{\rmd\ }{\rmd y};&
  \rmd\uir{X_2}{\myh_2\myhbar_4} &= 2\pi\rmi \myh_2+ \rmi\pi \myhbar_4 y^2;\\
  \nonumber 
  \rmd\uir{X_3}{\myh_2\myhbar_4} &=- 2\pi\rmi \myhbar_4 y;&
  \rmd\uir{X_4}{\myh_2\myhbar_4}&= 2\pi\rmi \myhbar_4\, I .
\end{align}

Recall~\cite{Kisil17a},  the \emph{lifting}
\(\FSpace{L}{\phi}(\Space{G}{}/Z) \rightarrow
\FSpace{L}{\myhbar_4}(\Space{G}{})\) to the space of functions \(F\)
having the property
\begin{equation}
  \label{eq:covariance-shear}
  F(x_1,x_2,x_3,x_4+z)=\rme^{- 2\pi\rmi \myhbar_4 z} F(x_1,x_2,x_3,x_4)\,.
\end{equation}
Then, \emph{Lie derivative} (left invariant vector fields)
\(\linv{X}\) for an element \(X\) of a Lie algebra \(\algebra{g}\) is
computed through the derived right regular representation of the
lifted function:
\begin{equation}
\label{Lie derivative def}
[\linv{X}F](g)=\frac{\rmd}{\rmd t}F(g\exp{tX})\bigg|_{t=0}
\end{equation}
for any differentiable function \(F\) on \(G\). 
For the group \(\Space{G}{}\) and functions
satisfying~\eqref{eq:covariance-shear} we find that:
  \begin{align}
  \label{eq:Lie-derivatives-A}
 \linv {X_1} &= \partial_{1};&
 \linv {X_2}
  &= \partial_{2}+ x_1\partial_{3}
    -\rmi \pi \myhbar_4  x_1^{2} I;\\
    \nonumber 
 \linv {X_3} &= \partial_{3}-{2 \pi \rmi} \myhbar_4 x_1 I ;&
 \linv {X_4} &= -{2\pi  \rmi}  \myhbar_4 I\,.
\end{align}

A direct calculation shows that all operator sets~\eqref{derived re
  reducibl re anharmo group},~\eqref{deri rep of irre shear group} and
\eqref{eq:Lie-derivatives-A} represent the Lie algebra \(\algebra{g}\)
of the group \(\Space{G}{}\).

\subsection{The group \(\Space{G}{}\), the Schr\"odinger group and symplectomorphisms}
\label{sec:shear-group-schr}

As was mentioned above, the Heisenberg group is a subgroup of the
group \(\Space{G}{}\). In its turn, the group \(\Space{G}{}\) is
isomorphic to a subgroup of the \emph{Schr\"odinger group}
\(\Space{S}{}\)---the group of
symmetries of the Schr\"odinger
equation~\cites{Niederer72a,KalninsMiller74a}, the harmonic
oscillator~\cite{Niederer73a}, other parabolic
equations~\cite{Wolf76a} and paraxial
beams~\cite{ATorre09a}. \(\Space{S}{}\) is also known as \emph{Jacobi
  group}~\cite{Berndt07a}*{\S\,8.5} due to its role in the theory of
Jacobi theta functions. A study of those important topics is outside
of scope of the present paper and we briefly recall a relevant part.
The Schr\"odinger group is the semi-direct product
\(\Space{S}{}=\Space{H}{}\rtimes_{A} \SL\), where \(\SL\) is the group
of all \(2\times 2\) real matrices with the unit determinant. The
action \(A\) of \( \SL\) on \(\Space{H}{}\) is given by
\citelist{\cite{deGosson11a}*{Ch.~4}
  \cite{deGosson06a}*{Ch.~7}\cite{Folland89}*{\S\,4.1}}:
\begin{equation}
  \label{action of sL2 on H}
  A(g): (x,y,s) \mapsto (ax+by,cx+dy,s),
\end{equation}
where
\(g=
\begin{pmatrix}
  a&b\\c&d
\end{pmatrix}\in\SL\) and  \((x,y,s)\in\Space{H}{}\).
Let
\begin{displaymath}
  N=\left\{ \begin{pmatrix}1&0\\x&1
\end{pmatrix},\  x\in \Space{R}{}\right\}  
\end{displaymath}
be the subgroup of \(\SL\), it is easy to check that \(\Space{G}{}\)
is isomorphic to the subgroup \(\Space{H}{} \rtimes_{A} N\) of
\(\Space{S}{}\) through the map in the exponential coordinates:
\begin{displaymath}
  (x_1,x_2,x_3,x_4)\mapsto  \left((x_1,x_3,x_4),n(x_2)\right)\in\Space{H}{} \rtimes_{A} N, 
\end{displaymath}
 where 
\((x_1,x_3,x_4)\in\Space{H}{}\) and \(n(x_2) :=\begin{pmatrix}1&0\\-x_2&1
\end{pmatrix}\in\SL\).

\begin{figure}[htbp]
  \centering
  \includegraphics[scale=.9]{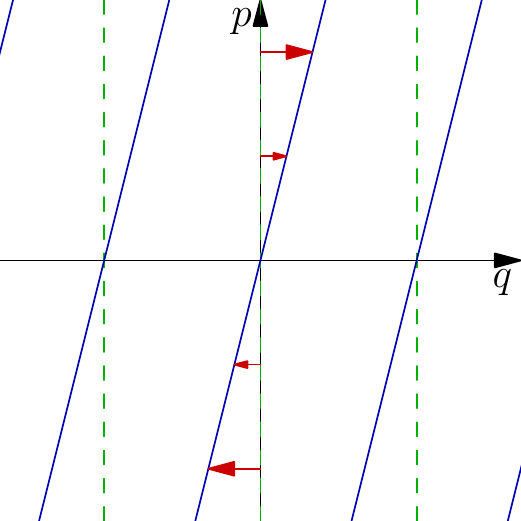}\qquad \qquad
  \includegraphics[scale=.9]{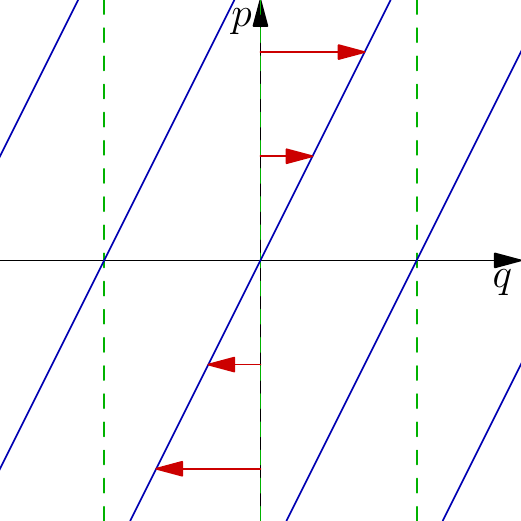}
  \caption[Shear transforms]{Shear transforms: dashed (green) vertical
    lines are transformed to solid (blue) slanted ones. An interpretation as
    a dynamics of a free particle: the momentum is constant, the
    coordinate is changed by an amount proportional to the momentum
    (cf. different arrows on the same picture) and the elapsed time
    (cf. the left and right pictures).}
  \label{fig:shear-transform}
\end{figure}
The geometrical meaning of this transformation is \emph{shear
  transform} with the angle \(\tan^{-1} x_2\), see Fig.~\ref{fig:shear-transform}:
\begin{equation}
  \label{eq:time-evolution-free-particle}
  n(x_2)(x_1,x_3):=\begin{pmatrix}1&0\\-x_2&1
  \end{pmatrix}
  \begin{pmatrix}x_1\\x_3\end{pmatrix}=\begin{pmatrix}x_1\\-x_2x_1+x_3\end{pmatrix}.
\end{equation}
Note, that  this also describes a physical picture for a particle
with coordinate \(x_3\) and the constant velocity \(x_1\): after a
period of time \(-x_2\) the particle will still have the velocity
\(x_1\) but its new coordinate will be \(x_3-x_2x_1\).  We shall refer
to both geometric and physical interpretation of the shear transform
in \S\,\ref{sec:harm-oscill-from-shear} in connection with the dynamic
of the harmonic oscillator.

Another important group of symplectomorphisms---squeezing---are
produced by matrices   \(\begin{pmatrix}
  a&0\\0&a^{-1}
\end{pmatrix}\in
\SL\). These transformations act transitively on the set of minimal
uncertainty states \(\phi\), such that \(\Delta_\phi Q \cdot
\Delta_\phi P= \frac{\myhbar}{2}\)---the minimal value admitted by the
Heisenberg--Kennard uncertainity relation~\cite{Folland89}*{\S\,1.3}.  

A related origin of the group \(\Space{G}{}\) is the universal enveloping algebra
\(\mathcal{H}\) of \(\algebra{h}\) spanned by elements \(Q\), \(P\)
and \(I\) with \([P,Q]=I\). It is known that the Lie algebra of Schr\"odinger group can
be identified with the subalgebra spanned by the elements
\(\{Q,P,I,Q^2, P^2,\frac{1}{2}(QP+PQ)\}\subset \mathcal{H}\). This
algebra is known as \emph{quadratic algebra} in quantum
mechanics~\cite{Gazeau09a}*{\S\,2.2.4}. From the above discussion
of the Schr\"odinger group, the identification
\begin{displaymath}
  X_1\mapsto P, \quad
  X_2\mapsto \textstyle \frac{1}{2}Q^2,\quad
  X_3\mapsto Q, \quad
  X_4\mapsto I
\end{displaymath}
embeds the Lie algebra \(\algebra{g}\) into \(\mathcal{H}\).  In
particular, the identification
\(X_2\mapsto \textstyle \frac{1}{2}Q^2\) was used in physical
literature to treat anharmonic oscillator with quartic
potential~\cites{AllenAnastassiouKlink97,Klink94,JorgensenKlink85}.
Furthermore, the group \(\Space{G}{}\) is isomorphic to the \emph{Galilei
group} via the identification of respective Lie algebras
\begin{displaymath}
  X_1\mapsto -Q, \quad
  X_2\mapsto \textstyle \frac{1}{2}P^2,\quad
  X_3\mapsto P, \quad
  X_4\mapsto I.
\end{displaymath}
That is, both algebras are related by Fourier transform.

We shall note that consideration of \(\Space{G}{}\) as a subgroup of
the Schr\"odinger group or the universal enveloping algebra
\(\mathcal{H}\) has a limited scope since only representations
\(\uir{}{\myh_2 \myhbar_4}\) with \(\myh_2=0\) appear as
restrictions of representations of Schr\"odinger group,
see~\citelist{\cite{deGosson11a}*{Ch.~7} \cite{deGosson06a}*{Ch.~7}
  \cite{Folland89}*{\S\,4.2}}.

\section{The induced coherent state transform and its image}
\label{sec:induc-wavel-transf}
We consider here the transformation which plays the crucial role in
geometrisation of many concepts. If this transformation is reduced
from the group \(\Space{G}{}\) to the Heisenberg group it coincides with the
Fock--Segal--Bargmann transform.

\subsection{Induced coherent state transform of the group $\Space{G}{}$}
\label{sec:induc-wavel-transf-shear}

Let \(G\) be a Lie group with a left Haar measure \(\rmd g\) and
\(\uir{}{}\) a unitary irreducible representation UIR of the group
\(G\) in a Hilbert space \(\FSpace{H}{}\). Then, we define the
coherent state transform as follows
\begin{defn} \cites{AliAntGaz14a,Kisil11c}
\label{orginal def of wavelet transfrom}
For a fixed vector \(\phi\in \FSpace{H}{}, \) called a
\emph{fiducial vector} (aka \emph{vacuum vector}, \emph{ground state},
\emph{mother wavelet}), the coherent state transform, denoted \(\oper{W}_{\phi}^{\uir{}{}}
\) (or just \(\oper{W}_{\phi}
\) when it is clear from the context),  of a vector \(f\in \FSpace{H}{}\) is given by:
\begin{equation*}
[\oper{W}_{\phi}^{\uir{}{}}f](g)=\scalar{\uir{}{}(g^{-1})f}{\phi}=\scalar{f}{\uir{}{}(g)\phi},\qquad g\in G.
\end{equation*}
\end{defn}

Let a fiducial vector \(\phi\) be a joint eigenvector of
\(\uir{}{}(h)\) for all \(h\) in a subgroup \(H\), that is
\begin{equation}
  \label{H induced wavelet proprety}
  \uir{}{}(h)\phi=\chi(h) \phi \quad \text{for all} \quad h\in H,
\end{equation}
where \(\chi\) is a character of \(H\). Then, straightforward
transformations show that~\citelist{\cite{Kisil11c}*{\S\,5.1}
  \cite{Kisil17a}*{\S\,5.1}}:
\begin{align}
\label{covriance property of wavelet}
[\oper{W}_{\phi}^{\uir{}{}}f](gh)
&=\overline{\chi}(h)[\oper{W}_{\phi}^{\uir{}{}}f](g).
\end{align}
This, in turn, indicates that the coherent state transform is entirely
defined via its values indexed by points of \(X=G/H\). This motivates
the following definition of the coherent state transform on the homogeneous space \(X=G/H:\)
 \begin{defn}
 \label{induced wavelet}
 For a group \(G\), its subgroup \(H\), a section
 \(\map{s}: G/H \rightarrow G\), UIR \(\uir{}{}\) of \(G\) in a
 Hilbert space \(\FSpace{H}{}\) and a vacuum vector \(\phi\)
 satisfying~\eqref{H induced wavelet proprety}, we define the
 \emph{induced coherent state transform} \(\oper{W}_{\phi}^{\uir{}{}}\) from
 \(\FSpace{H}{}\) to a space of function \(\FSpace{L}{\phi}(G/H)\) by the
 formula
 \begin{equation}
 \label{induced wavelet transform}
[\oper{W}_{\phi}^{\uir{}{}} f](x)=\langle f,\uir{}{}(\mathsf{s}(x))\phi\rangle,\qquad x\in G/H\,.
 \end{equation}
 \end{defn}

  \begin{prop}
  \citelist{\cite{Kisil11c}*{\S\,5.1}
  \cite{Kisil17a}*{\S\,5.1}}
    \label{intertwing repre} Let \(G\), \(H\), \(\uir{}{}\), \(\phi\)
    and \(\oper{W}_{\phi}^{\uir{}{}}\) be as in
    Definition~\ref{induced wavelet} and \(\chi\) be a character
    from~\eqref{H induced wavelet proprety}. Then the induced coherent state
    transform intertwines \(\uir{}{}\) and \(\tilde{\uir{}{}}\)
  \begin{equation}
  \label{invariance of wavelet image}
  \oper{W}^{\uir{}{}}_{\phi}\uir{}{}(g)=\tilde{\uir{}{}}(g)\oper{W}^{\uir{}{}}_{\phi},
  \end{equation}
  where \(\tilde{\uir{}{}}\) is an induced representation from the
  character \(\chi\) of the subgroup \(H\).

  In particular,~\eqref{invariance of wavelet image} means that the
  image space \(\FSpace{L}{\phi}(G/H)\) of the induced coherent state transform is
  invariant under \(\tilde{\uir{}{}}\).
\end{prop}

For the subgroup \(H\) being the centre
\(Z=\{(0,0,0,x_4)\in\Space{G}{}: x_4\in \Space{R}{}\}\) of
\(\Space{G}{}\), the representation
\(\uir{}{\myh_2\myhbar_4}\)~\eqref{irreducible repre of A} and the
character \(\chi(0,0,0,x_4)=\rme^{2\pi\rmi \myhbar_4 x_4}\) any
function \(\phi\in\FSpace{L}{2}(\Space{R}{})\) satisfies the
eigenvector property~\eqref{H induced wavelet proprety}.  Thus, for
the respective homogeneous space \(\Space{G}{}/Z\sim \Space{R}{3}\)
and the section
\(\mathsf{s}:\Space{G}{}/Z\to \Space{G}{}; \
\mathsf{s}(x_{1},x_{2},x_{3})=(x_{1},x_{2},x_{3},0)\) the induced
coherent state transform is:
 \begin{align}
\nonumber
   [\oper{W}_{\phi}f](x_{1},x_{2},x_{3})&=\langle f,\uir{}{\myh_2\myhbar_4}(\mathsf{s}(x_{1},x_{2},x_{3}))\phi\rangle\\
 \nonumber
 &=\langle f,\uir{}{\myh_2\myhbar_4}(x_{1},x_{2},x_{3},0)\phi\rangle\\
 \nonumber
 &=\int_{\Space{R}{}} f(y)\, \overline{\uir{}{\myh_2\myhbar_4}(x_{1},x_{2},x_{3},0)\,\phi(y)}\,\rmd y
 \\
 \nonumber
 &=\int_{\Space{R}{}} f(y)\,\rme^{-2\pi \rmi (\myh_2x_2+\myhbar_4(-x_3y+\frac{1}{2}x_2y^2))}\,
 \overline{\phi}(y-x_1)\,\rmd y\\
\label{eq:wavelet-transform-shear-group}
 &=\rme^{-2\pi \rmi \myh_2x_2}\int_{\Space{R}{}} f(y)\,\rme^{-2\pi \rmi \myhbar_4(-x_3y+\frac{1}{2}x_2y^2)}\,
 \overline{\phi}(y-x_1)\,\rmd y.
 \end{align}
 The last integral is a composition of the following
 three unitary operators of \(\FSpace{L}{2}(\Space{R}{2})\):
 \begin{enumerate}
 \item The measure-preserving change of variables
   \begin{equation}
     \label{eq:T-change-variables}
     T:F(x_1,y)\mapsto F(y,y-x_1)\,,
   \end{equation}
   where \(F(x_1,y):=(f\otimes
   \overline{\phi})(x_1,y)=f(x_1) \overline{\phi}(y)\), that is, \(F\) is defined on the
   tenser product  \(\FSpace{L}{2}(\Space{R}{})\otimes \FSpace{L}{2}(\Space{R}{})\) which is
   isomorphic to \(\FSpace{L}{2}(\Space{R}{2})\);
   \item the operator of multiplication by an unimodular  function
     \(\psi_{x_2}(x_1,y)=\rme^{- \pi{\rmi} \myhbar_4 x_2 y^2}\)
     \begin{equation}
       \label{eq:M-multiplic-unimodular}
       M_{x_2}: F(x_1,y)\mapsto \rme^{-\pi{\rmi} \myhbar_4 x_2 y^2}F(x_1,y),\quad  x_2\in \Space{R}{};
     \end{equation}
   \item and the partial inverse Fourier transform in the second variable
     \begin{equation}
       \label{eq:F-Fourier-partial}
       [\oper{F}_{2}F](x_1,x_3)=\int_{\Space{R}{}}F(x_1,y)\rme^{2\pi\rmi \myhbar_4 y x_3}\,\rmd y.
     \end{equation}
 \end{enumerate}

 Thus, [\(\oper{W}_{\phi}f](x_1,x_2,x_3) =\rme^{-2\pi \rmi \myh_2
   x_2}[\oper{F}_{2}\circ M_{x_2}\circ T]F(x_1,x_3)\) and we obtain
 \begin{prop}
   \label{prop:wavelet-is-isometry}
   For a fixed \(x_2 \in \Space{R}{}\), the map
   \(f \otimes \overline{\phi} \mapsto [\oper{W}_{\phi}f](\cdot,x_2,\cdot)\)
   is a unitary operator on \(\FSpace{L}{2}(\Space{R}{2})\).
 \end{prop}
  Such an induced coherent state transform also respects the Schwartz space, that is, if
 \(f,\phi \in \FSpace{S}{}(\Space{R}{})\) then
 \([\oper{W}_{\phi}f](\cdot,x_2,\cdot)\in
 \FSpace{S}{}(\Space{R}{2})\). This is because
 \(\FSpace{S}{}(\Space{R}{2})\) is
 invariant under each
 operator~\eqref{eq:T-change-variables}--\eqref{eq:F-Fourier-partial}.

 \begin{rem}
   \label{re:square-integrability}
   It is known fact \cite{CorwinGreenleaf90a}*{\S\,4.5} that the
   coherent state transform on a nilpotent Lie group, cf.
   Definition~\ref{orginal def of wavelet transfrom}, does not produce
   an \(\FSpace{L}{2}\)-function on the entire group but may rather do
   on a certain homogeneous space. The leading example is the
   Heisenberg group when considering the homogeneous space
   \(\Space{H}{}/Z\). In the context of coherent state transform, two
   types of modified square-integrability are
   considered~\cite{CorwinGreenleaf90a}*{\S\,4.5}: modulo the group's
   center and modulo the kernel of the representation. The first
   notion is not applicable to the group \(\Space{G}{}\): the coherent
   state transform~\eqref{eq:wavelet-transform-shear-group} does not
   define a square-integrable function on
   \(\Space{G}{}/Z\sim \Space{R}{3}\) or a larger space
   \(\Space{G}{}/\mathrm{ker}\uir{}{\myh_2 \myhbar_4}\).  On the other
   hand, the representation \(\uir{}{\myh_2 \myhbar_4}\) is
   square-integrable modulo the subgroup
   \(H=\{(0,x_2,0,x_4)\in \Space{G}{}: x_2,x_4\in
   \Space{R}{}\}\). However, the theory of
   \(\alpha\)-admissibility~\cite{AliAntGaz14a}*{\S\,8.4}, which is
   supposed to work for such cases, reduces the consideration to the
   Heisenberg group since \(\Space{G}{}/H \sim \Space{H}{}/Z\).  It
   shall be seen later~\eqref{solution to the shrodinger equation for
     shear case} that the action of \((0,x_2,0,0)\in H\) will be
   involved in important physical and geometrical aspects of the state
   of the harmonic oscillator and shall not be factored out.
   Our study provides an example of the theory of wavelet transform with
   non-admissible mother wavelets~\cites{Kisil09d,Kisil10c,Kisil98a,Zimmermann06,Guerrero18}.
\end{rem}
In view of the mentioned above insufficiency of square integrability
modulo the subgroup
\(H=\{(0,x_2,0,x_4)\in \Space{G}{}: x_2,x_4\in \Space{R}{}\}\), we
make the following
\begin{defn}
  For a fixed unit vector \(\phi\in\FSpace{L}{2}(\Space{R}{})\), let
  \(\FSpace{L}{\phi}(\Space{G}{}/Z)\) denote the image space of the
  coherent state transform
  \(\oper{W}_{\phi}\)~\eqref{eq:wavelet-transform-shear-group}
  equipped with the family of inner products parametrised by \(x_2 \in \Space{R}{}\)
  \begin{equation}
    \label{inner product of the image}
    \scalar[x_2]{u}{v} :=\int_{\Space{R}{2}}
    u(x_1,x_2,x_3)\,\overline{v(x_1,x_2,x_3)}\,\myhbar_4\,\rmd{x_1}\rmd{x_3}\,.
  \end{equation}
  The respective norm is denoted by \(\norm[x_2]{u}\).   
\end{defn}
\begin{rem}
\label{physical units}
In physical applications the elements \(x_1\), \(x_2\) and \(x_3\) are
physical quantities and shall have physical units, cf. the
consideration in~\cites{Kisil02e,Kisil17a} 
 for the Heisenberg group
and quantization problem in general. Let \(L\) denote the unit of
length, \(M\) that of mass and \(T\) that of time. Then, \(x_1\) has
dimension \(\frac{T}{ML}\) (reciprocal to momentum) and \(x_3\) has
dimension \(1/L\) (reciprocal to position.)  So, from~\eqref{anh
  comutation relation} \(x_2\) has dimension \(\frac{M}{T}\). Also,
\(\myh_2\), the dual to \(x_2\) has reciprocal dimension to
\(x_2\), that is, it has dimension \(\frac{T}{M}\) as well as
\(\myhbar_4\) has dimension \(\frac{ML^2}{T}\) of action which is
reciprocal to the dimension of the product \(x_1x_3\) or \(x_4\).  Dimensionality
of \(\myhbar_4\) coincides with that of Planck constant. Thus the
factor \(\myhbar_4\) in the measure \(\myhbar_4\,\rmd{x_1}\rmd{x_3}\)
makes it dimensionless, which is a natural physical
requirement.  Note also that the dimensionality of \(\rmd\uir{X_j}{}\) is reciprocal to that of \(x_j.\)
\end{rem}
It follows from Prop.~\ref{prop:wavelet-is-isometry} that
\(\norm[x_2]{u}=\norm[x'_2]{u}\) for any \(x_2\),
\(x_2'\in\Space{R}{}\) and \(u\in \FSpace{L}{\phi}(\Space{G}{}/Z)\).
In the usual way \cite{Folland89}*{(1.42)}
 the isometry from
Prop.~\ref{prop:wavelet-is-isometry} implies the following
\emph{orthogonality relation}.
\begin{cor}
  Let \(f_1\), \(f_2\), \(\phi_1\),
  \(\phi_2\in\FSpace{L}{2}(\Space{R}{})\) then:
  \begin{equation}
    \label{eq:anharm-othog-relation}
    \scalar[x_2]{\oper{W}_{\phi_1}f_1}{\oper{W}_{\phi_2}f_2}=
  \scalar{f_1}{f_2} \overline{\scalar{\phi_1}{\phi_2}} \qquad \text{ for any
  } x_2\in \Space{R}{}\,.
  \end{equation}
\end{cor}
\begin{cor}
  Let \(\phi\in\FSpace{L}{2}(\Space{R}{})\) have unit norm, then the
  coherent state transform \(\oper{W}_{\phi}\) is an isometry with respect
  to the inner product~\eqref{inner product of the image}.
\end{cor}
\begin{proof}
  It is an immediate consequence of the previous
  corollary. Alternatively,  for \(f\in\FSpace{L}{2}(\Space{R}{})\):
  \begin{displaymath}
    \norm[\FSpace{L}{2}(\Space{R}{})]{f}=
    \norm[\FSpace{L}{2}(\Space{R}{2})]{f\otimes \phi}
    =\norm[x_2]{\oper{W}_{\phi} f}\,,
  \end{displaymath}
  as follows from the isometry
  \(\oper{W}_{\phi}: \FSpace{L}{2}(\Space{R}{}) \rightarrow
  \FSpace{L}{2}(\Space{R}{2})\) in Prop.~\ref{prop:wavelet-is-isometry}.
\end{proof}

 An inverse of the unitary operator \(\oper{W}_{\phi}\) is given
 by its adjoint \(\oper{M}_{\phi}(x_2)=\oper{W}_{\phi}^*\) with
 respect to the inner product~\eqref{inner product of the image} parametrised by
 \(x_2\):
 \begin{equation}
\label{contravariant transform}
[\oper{M}_{\phi}(x_2)f](t)=\int_{\Space{R}{2}}f(x_1,x_2,x_3)
\uir{}{}(x_1,x_2,x_3,0) \phi(t)\, \myhbar_4\, \rmd x_1\,\rmd x_3.
\end{equation}
More generally, for an analysing vector \(\phi\) and a  reconstructing
vector \(\psi\),  for any \(f\), \( g\in \FSpace{L}{2}(\Space{R}{})\)
the orthogonality condition~\eqref{eq:anharm-othog-relation} implies:
\begin{align*}
  \scalar{\oper{M}_{\psi}(x_2)\circ \oper{W}_{\phi}f}{g}
  &= \scalar[x_2]{\oper{W}_{\phi}f}{\oper{W}_{\psi}g}\\
  &= \scalar{f}{g} \scalar{\psi}{\phi}\,.
\end{align*}
Then \(\oper{M}_{\psi}(x_2)\circ \oper{W}_{\phi}=
\scalar{\psi}{\phi} I\) and if \(\scalar{\psi}{\phi} \neq0\) then
\(\oper{M}_{\psi}(x_2)\) is a left inverse of
\(\oper{W}_{\phi}\) up to a factor.

By Prop.~\ref{intertwing repre}  the transform \(\oper{W}_{\phi}\) intertwines \(\uir{}{\myh_2\myhbar_4}\) with the following representation on \(\FSpace{L}{2}(\Space{R}{3})\) (see~\eqref{reducible re of anhar def})
\begin{multline}
 \nonumber
  \label{reducible rep shear group}
  [\uir[2]{}{\myh_4}(y_1,y_2,y_3,y_4)f](x_1,x_2,x_3)= \rme^{ 2\pi\rmi \myhbar_4(y_4-y_1y_3+\frac{1}{2}y_1^2y_2+y_1x_3-\frac{1}{2}y_1^2 x_2)}\\
\times f(x_1-y_1,x_2-y_2,x_3-y_1x_2+y_1y_2-y_3)\,.
\end{multline}
We consider this representation restricted to the image space
\(\FSpace{L}{\phi}(\Space{G}{}/Z)\), unitarity of
\(\uir[2]{}{\myhbar_4}\) is easily seen.

\subsection{Characterisation of the image space}
\label{sec:char-image-space}
The following result plays a fundamental role in exploring the nature
of the image space of the coherent state transform and is a recurrent theme of
our investigation, see also \citelist{\cite{Kisil11c}*{\S\,5}\cite{Kisil13c} \cite{Kisil17a}*{\S\,5.3}}.
\begin{cor}[Analyticity of the coherent state transform,~\cite{Kisil11c}*{\S\,5}] 
  \label{co:cauchy-riemann-integ}
  Let \(G\) be a group and \(\rmd g\) be a measure on \(G\). Let \(\uir{}{}\) be a unitary representation of \(G\), which can be
  extended by integration to a vector space \(V\) of functions or
  distributions on \(G\).  Let a fiducial vector \(\phi\in \FSpace{H}{}\) satisfy the
  equation
  \begin{equation}
  \label{auxiliary condition for fiducial }
    \int_{G} d(g)\, \uir{}{}(g) \phi\,\rmd g=0,
  \end{equation}
  for a fixed distribution \(d(g) \in V\). Then,  any coherent state transform
  \(\tilde{v}(g)=\scalar{v}{\uir{}{}(g)\phi}\) obeys the condition:
  \begin{equation}
     \label{eq:dirac-op}
     D\tilde{v}=0,\qquad \text{where} \quad D=\int_{G} \bar{d}(g)\, R(g) \,\rmd g\,,
  \end{equation}
  with \(R\) being the right regular representation of \(G\) and
  \(\bar{d}(g)\) is the complex conjugation of \({d}(g)\).
\end{cor}

\begin{example}

To describe the image space \(\FSpace{L}{\phi}(\Space{G}{}/Z)\) of the respective induced  coherent state transform, we employ Corollary~\ref{co:cauchy-riemann-integ}. This requires a particular
choice of a fiducial vector \(\phi\) such that  \(\phi\) lies in
\(\FSpace{L}{2}(\Space{R}{})\) and  \(\phi\) is a null solution of an
operator of the form \eqref{auxiliary condition for fiducial }. For
simplicity, we consider the following first-order operator, which
represents an element of the Lie algebra \(\algebra{g}\), cf.~\eqref{deri rep of irre shear group}:
\begin{align}
  \nonumber 
  \rmd\uir{\rmi X_{1}+\rmi aX_2 + E X_3}{\myh_2\myhbar_4}
  &= \rmi
    \rmd\uir{X_{1}}{\myh_2\myhbar_4}
    +\rmi a
    \,\rmd\uir{X_{2}}{\myh_2\myhbar_4}
    +E\,\rmd\uir{X_{3}}{\myh_2\myhbar_4}\\
  \label{combination of deri rep for irr rep anhar}
  &=-\rmi \frac{\rmd\ }{\rmd y}     - \pi   \myhbar_4
    a y^2-2\pi\rmi E\myhbar_4 y  -2\pi\myh_2 a 
    \,,
\end{align}
where \(a\) and \(E\) are some real constants. It is clear that, the function
 \begin{equation}
   \label{mother-wavelet for kirilov re}
   \phi(y)=c\exp\left(\frac{ \pi\rmi a\myhbar_{4} }{3} y^{3}-  \pi E\myhbar_{4}y^{2}+ 2\pi\rmi a\myh_{2}y\right),
\end{equation}
is a generic solution of \eqref{combination of deri rep for irr rep anhar} where \(c\) is an arbitrary constant determined by \(\FSpace{L}{}_2\)-normalisation while square integrability of \(\phi\) requires
that \(E\myhbar_4\) is strictly positive. Furthermore, it is
sufficient for the purpose of this work to use the simpler fiducial
vector corresponding to the value\footnote{The case of \(a\neq 0\)
  would correspond to Airy beams~\cite{ATorre09a}.}
\(a=0\), thus we set
\begin{equation}
  \label{the right choise of mother wavelet}
  \phi(y)=c\,\rme^{-\pi E \myhbar_4 y^2},\qquad \myhbar_4>0,\  E>0.
\end{equation}

Since the function \(\phi(y)\)~\eqref{the right choise of mother
  wavelet} is a null-solution of the operator~\eqref{combination of deri rep for irr rep anhar} with \(a=0\), the image space
\(\FSpace{L}{\phi}(\Space{G}{}/Z)\) can be described through the
respective derived right regular representation (Lie
derivatives)~\eqref{Lie derivative def}.  Specifically, 
Corollary~\ref{co:cauchy-riemann-integ} with the distribution
\begin{displaymath}
  d(x_1,x_2,x_3,x_4)= \rmi\,  \delta'_1(x_1,x_2,x_3,x_4)
  + E\,   \delta'_3(x_1,x_2,x_3,x_4)\,,
\end{displaymath}
where \(\delta'_j\) is the partial derivative of the Dirac
delta distribution \(\delta\) with respect to \(x_j\), matches~\eqref{combination of deri
  rep for irr rep anhar}. Thus, any function \(f\) in
\(\FSpace{L}{\phi}(\Space{G}{}/Z)\) for \(\phi\)~\eqref{the right
  choise of mother wavelet} satisfies \(\oper{C} f =0\) for the
partial differential operator: 
\begin{align}
  \label{eq:analyticity-anh}
  \oper{C}
  =\left( -\rmi\linv {X_1}+E\linv {X_3}\right) 
  = -\rmi\partial_{1}+ E \partial_{3} -2\pi\rmi  \myhbar_4 E { x_1 }\, ,
\end{align}
where Lie derivatives \eqref{eq:Lie-derivatives-A} are used.


Due to the explicit similarity to the Heisenberg group case
with the Cauchy--Riemann equation, we call~\eqref{eq:analyticity-anh}
the analyticity condition for the coherent state transform.
 Indeed, it can be easily verified that 
\begin{align*}
[\mathcal{W}_{\phi}k](x_1,x_2,x_3)=\exp(- 2\pi\rmi \myh_2 x_2+\pi{\rmi} \myhbar_4 x_1x_3-\frac{ \pi\myhbar_4}{2E}(E^2x_1^{2}+x_3^{2}))B_{x_2}(x_1,x_3)
\end{align*} where 
\begin{align*}
B_{x_2}(x_1,x_3)&=\int_{\mathbb{R}} \rme^{-\pi{\rmi} \myhbar_4 x_2 y^2}k(y) \rme^{\frac{ \pi\myhbar_4}{2E} (x_3-\rmi Ex_1)^2+2\pi\rmi \myhbar_4 (x_3-\rmi Ex_1)y-\pi \myhbar_4E y^2}\;\rmd y\\
&= [(V_E \circ M_{x_2})k](x_1,x_3)
\end{align*} and $V_E$ is Fock--Segal--Bargmann type transform \cites{Folland89, Kisil11c, Kirillov04a, Neretin11a} 
 and \(M_{x_2}\) is a multiplication operator by \(\rme^{-\pi \myhbar_4 x_2 y^2}\). Thus, by condition~\eqref{eq:analyticity-anh} we have 
\begin{align*}
&[-\rmi\partial_{1}+E \partial_{3}\\
&\quad - 2\pi\rmi  \myhbar_4 E { x_1 }]\{\exp(- 2\pi\rmi \myh_2 x_2+\pi{\rmi} \myhbar_4 x_1x_3-\frac{\pi \myhbar_4}{2E}(E^2x_1^{2}+x_3^{2}))B_{x_2}(x_1,x_3)\}=0.
\end{align*} That is,
\begin{equation}
(-\rmi\partial_{1} +E \partial_{3})B_{x_2}(x_1,x_3)=0, 
\end{equation} which can be written as 
\begin{equation}
\partial_{\overline{z}} B_{x_2}(z)=0, 
\end{equation}
where \(z=x_3-\rmi Ex_1\) and \(\partial_{\overline{z}}=\frac{1}{2}(\frac{-\rmi}{E}\partial_{1}+ \partial_{3})\)---a Cauchy--Riemann type operator. Thus, \(B_{x_2}(z)\) is entire on the complex plane \(\Space{C}{}\) parametrised by points \((x_1,x_3)\). As such, the coherent state transform gives rise to the space consisting of analytic functions \(B_{x_2}(z)\) which are square-integrable with respect to the measure \(\rme^{-\frac{ \pi\myhbar_4}{E}\modulus {z}^{2}} \rmd z\).
\end{example}

A notable difference between the group \(\Space{G}{}\) and the Heisenberg
group is the presence of an additional second-order condition, which is
satisfied by any function \(f\in \FSpace{L}{\phi}(\Space{G}{}/Z)\) for
any fiducial vector \(\phi\). Indeed, the specific structure of the
representation~\eqref{deri rep of irre shear group} implies that any function
\(\phi\) satisfies the relation for the derived representation,
\begin{align}
   \label{eq:Casimir-fixed-value-action}
 \left((\rmd\uir{X_3}{\myh_2\myhbar_4})^2
 -2\,\rmd\uir{X_2}{\myh_2\myhbar_4} \rmd\uir{X_4}{\myh_2\myhbar_4}- {8\pi^2 \myh_2\myhbar_4}
   I\right)\phi =0.
\end{align}
This can by verified by~\eqref{deri rep of irre shear group} directly, but a deeper insight follows from an
interpretation of~\eqref{eq:Casimir-fixed-value-action} as the
statement that the Casimir operator \(C=X_3^2-2X_1X_2\)
\ acts on the irreducible component as multiplication by the scalar
\({8\pi^2 \myh_2\myhbar_4}\). Also Casimir operator \(C\) appears in the
Kirillov orbit method~\citelist{\cite{CorwinGreenleaf90a}*{Ex.~3.3.9}
  \cite{Kirillov04a}*{\S\,3.3.1}} in the natural parametrisation of
coadjoint orbits---the topic further developed in~\cites{AldayaNavarro-Salas91a,AldayaGuerrero01a}.

From here we can proceed in either way:
\begin{enumerate}
\item Corollary~\ref{co:cauchy-riemann-integ} with the distribution
  \begin{displaymath}
    d(x_1,x_2,x_3,x_4)=  \delta_{33}^{(2)}(x_1,x_2,x_3,x_4)
    -2\, \delta'_1(x_1,x_2,x_3,x_4) \cdot \delta'_2(x_1,x_2,x_3,x_4)\,,
  \end{displaymath}
   asserts that the image
   \(f\in \FSpace{L}{\phi}(\Space{G}{}/Z)\) of the coherent state transform
   \(\oper{W}_{\phi}\) is annihilated by the respective Lie derivatives
   operator  \(\oper{S}f=0\) where, cf.~\eqref{eq:Lie-derivatives-A}:
   \begin{align}
     \label{eq:general-annih}
     \oper{S}&=(\linv {X_3})^2-2 \linv {X_2}\linv {X_4} - {8 \pi^2\myh_2\myhbar_4} I\\
     \nonumber
             &=  \partial_{33}^{2}
               +{4\pi \rmi}  \myhbar_4 \partial_{2}
               -8\pi^2  \myh_2  \myhbar_4 I\,.
   \end{align}
 \item The representation, cf.~\eqref{derived re reducibl re anharmo group}:
   \begin{equation}
     \rmd\uir[2]{C}{\myhbar_4}=(\rmd\uir[2]{X_3}{\myhbar_4})^2-\rmd\uir[2]{X_2}{\myhbar_4} \rmd\uir[2]{X_4}{\myhbar_4}
   \end{equation}
   of the Casimir operator \(C\) takes the constant value
   \({8 \pi^2\myh_2\myhbar_4}\) on
   \(\FSpace{L}{\phi}(\Space{G}{}/Z)\). Note, that this produces
   exactly~\eqref{eq:general-annih} because for the Casimir operator
   the left and the right actions coincide.
\end{enumerate}
The relation~\eqref{eq:general-annih} will be called the
\emph{structural condition} because it is determined by the structure
of the Casimir operator.  Note,
that~\eqref{eq:general-annih} is a Schr\"odinger equation of a free
particle with the time-like parameter \(x_2\). Thus, the structural
condition is the quantised version of the classical
dynamics~\eqref{eq:time-evolution-free-particle} of a free particle
represented by the shear transform, see the discussion of this
after~\eqref{eq:time-evolution-free-particle}.

Summing up, the physical characterisation of the
\(\FSpace{L}{\phi}(\Space{G}{}/Z)\) is as follows:
\begin{enumerate}
\item The restriction of a function
  \(f\in \FSpace{L}{\phi}(\Space{G}{}/Z)\) to the plane \(x_2=0\) (a
  model of the phase space)
  coincides with the FSB image of the respective state,
\item The function \(f\) is a continuation from the plane \(x_2=0\) to
  \(\Space{R}{2}\times \Space{R}{}\) (the product of phase space and
  timeline) by free time-evolution of the quantum system.
\end{enumerate}
This physical interpretation once more explains the identity
\(\norm[x_2]{f}=\norm[x'_2]{f}\): the energy of a free state is
constant in time. 

Summarising analysis in this subsection we specify the ground states
and the respective coherent state transforms, which will be used below.
\begin{example}
Consider our  normalised  fiducial vector \(\phi_{E}(y)=({2\myh_2 E})^{1/4}\rme^{-\pi \myhbar_4E
  y^2}\)  with a parameter \(E>0\). It is
  normalised with respect to the norm 
  \(\norm{f}^2=\int_{\Space{R}{}}\modulus{f(y)}^{2}
  \sqrt{\frac{\myhbar_4}{\myh_2}}\;\rmd y\). Note, that for
  \(\phi_{E}(y)\) to be dimensionless, \(E\) has to be of  dimension
  \(M/T\), this follows from the fact that \(y\) has dimension
  \(T/ML\), thus for this reason we attach the factor
  \(\sqrt{\frac{\myhbar_4}{\myh_2}}\) to the measure so the measure is
  dimensionless. Then we calculate the coherent state transform of a
  minimal uncertainty state \(\phi_{q}(y)\) (\(q>0\)) as follows:
 \begin{align}
 \nonumber
[\mathcal{W}_{\phi_E}\phi_{q}](x_{1},x_{2},x_{3})&= \sqrt{2\myhbar_4}(qE)^{1/4}\int_\mathbb{R}\rme^{-\pi \myhbar_{4}q y^{2}}
 \rme^{-2\pi \rmi(h_2x_2+\myhbar_4(-x_3y+\frac{1}{2}x_2y^2))}\\
 \nonumber 
  &\qquad\times\rme^{-\pi \myhbar_{4}E(y-x_1)^{2}}\; \rmd y\\
  \nonumber
  &=\sqrt{2\myhbar_4}(qE)^{1/4}\rme^{-\pi \myhbar_{4}Ex_1^{2}-2\pi \rmi h_2x_2}\int_\mathbb{R} \rme^{-\pi \myhbar_4(\rmi x_2+E+q)y^2}\\
  \nonumber
  &\qquad\times \rme^{2\pi \myhbar_4(Ex_1+\rmi x_3)y}\;\rmd y
  \\ 
 \label{wavelet of mother wavelet shear group}
  &={ \sqrt{2}(qE)^{1/4}}\frac{\exp\left(-\pi h_4E x_1^2-2\pi \rmi h_2 x_2 - \pi \myhbar_4\frac{(-\rmi Ex_1+ x_3)^2}{\rmi x_2+E+q} \right)}{\sqrt{\rmi x_2+ E+q}}.
 \end{align}
Clearly, the function~\eqref{wavelet of mother wavelet shear group} is dimensionless and satisfies conditions \eqref{eq:analyticity-anh} and \eqref{eq:general-annih}. It shall be seen later that such a function is an eigenstate (vacuum state) of the harmonic oscillator Hamiltonian acting on the image space \(\FSpace{L}{\phi}(\Space{G}{}/Z)\) of the respective induced coherent state transform. Moreover, this function represents a minimal uncertinity state in the space \(\FSpace{L}{\phi}(\Space{G}{}/Z)\) for any \(x_2\in \mathbb{R}\) and any \(E>0\).
\end{example}

\section{Harmonic oscillator through reduction of order of a PDE}
\label{sec:harm-oscill-thro}
Here we systematically treat the harmonic oscillator with the method
of reducing PDE order for geometrisation of the dynamics. First we use
the Heisenberg group and find that geometrisation condition
completely determines which fiducial vector needs to be used. The
treatment of the group \(\Space{G}{}\) provides
 the wider opportunity: any
minimal uncertainty state can be used for the coherent state transform with
geometric dynamic in the result. 
In the next section we provide eigenfunctions
and respective ladder operators.
\subsection{Harmonic oscillator from the Heisenberg group}
\label{sec:harm-oscill-heisenberg}

Here we use a simpler case of the Heisenberg group to illustrate the
technique which will be used later.  The fundamental importance of the
harmonic oscillator stimulates exploration of different
approaches. Analysis based on the pair of ladder operators elegantly
produces the spectrum and the eigenvectors.  However, this technique is
essentially based on a particular structure of the Hamiltonian of
harmonic oscillator expressed in terms of the Weyl Lie algebra. Thus,
it loses its efficiency in other cases. In contrast, our method is
applicable for a large family of examples since it has a more general
nature.

The harmonic oscillator with constant mass \(m\) and frequency
\(\omega\) has the classic Hamiltonian
\(\frac{1}{2}(p^2/m+m \omega^2 q^2)\). Its Fock--Segal--Bargmann
quantisation, denoted by \(H\), acts on
\(\FSpace{L}{\phi}(\Space{H}{}/Z)\) parametrised by points \((x_1,x_3)\)
and is given by
\begin{align}
   \label{eq:harmonic-hamilt}
  Hf(x_1,x_3)&=\left(\frac{1}{2m}\left(\frac{\rmi\rmd\tilde{\sigma}_\myhbar^{X_1}}{\sqrt{2\pi}}\right)^2+\frac{m\omega^2}{2}\left(\frac{\rmi\rmd\tilde{\sigma}_\myhbar^{X_3}}{\sqrt{2\pi}}\right)^2\right)f(x_1,x_3)\\
  \nonumber
  &=
    -\frac{1}{4\pi m} \partial_{11}^{2}(f)(x_1,x_3)-\frac{m\omega^2}{4\pi}\partial_{33}^{2}(f)(x_1,x_3)\\
  \nonumber
   &\quad   +\frac{ \rmi \myhbar}{m} x_3
     \partial_{1}(f)(x_1,x_3) +  \frac{\pi \myhbar^{2}}{m} x_3^{2}f(x_1,x_3).
\end{align}

The time evolution of the harmonic oscillator is defined by the the Schr\"{o}dinger equation for  Hamiltonian \(H;\)
 \begin{equation}
 \label{hami for harminc on Heisenberg group}
 {\rmi \myhbar} \partial_tf(t;x_1,x_3) -Hf(t;x_1,x_3)=0
\end{equation}
for \(f(0,x_1,x_3)\) in
\(\FSpace{L}{\phi}(\Space{H}{}/Z)\). As was mentioned, the
creation-annihilation pair produces spectral
decomposition of \(H\). Our aim is to describe dynamics of an arbitrary
observable in geometric terms by lowering the order of the differential operator~\eqref{eq:harmonic-hamilt}.

Indeed, the structure of \(\FSpace{L}{\phi}(\Space{H}{}/Z)\) enables to make
a reduction to the order of equation~\eqref{eq:harmonic-hamilt} using
the analyticity condition.  For Gaussian \(\phi(y)=\rme^{-\pi \myhbar Ey^2}\) (\(E>0\)) we have (the annihilation operator)
\begin{displaymath}
 \rmi \rmd{\sigma_{\myhbar}^{X_1}}\phi(y)+E\rmd{\sigma_{\myhbar}^{X_3}}\phi(y)=0.\end{displaymath}
So, the respective image of induced coherent state
transform, \( f(x_1,x_3)=[\oper{W}^{\sigma_\myhbar}_{\phi}k](x_1,x_3)\)\quad  for \(k\in \FSpace{L}{2}(\Space{R}{})\), is annihilated by the operator (the analyticity condition):
\begin{align}
  \label{eq:cauchy-riemann}
 \oper{D}&=-\rmi \linv {X_1}+ E
\linv {X_3}\\
  \nonumber
  &=-\rmi\partial_{1}+ E  \partial_{3}-2\pi\rmi \myhbar
     E x_1  I\,,
\end{align}
we still write \(f(0,x_1,x_3)\) as \(f(x_1,x_3)\).
From the analyticity condition~\eqref{eq:cauchy-riemann} for induced coherent state transform,
operator \((A\partial_{1}+\rmi B
\partial_{3}+CI)(-\rmi\linv {X_1}+E \linv {X_3})\) vanishes on any
\(f\in \FSpace{L}{\phi}(\Space{H}{}/Z)\) for any \(A\), \(B\)
and \(C\). Thus, we can adjust the
Hamiltonian~\eqref{eq:harmonic-hamilt} by adding such a term
\begin{align*}
\tilde{H}&= H+(A\partial_{1}+\rmi B
  \partial_{3}+CI)(-\rmi\linv {X_1}+ E \linv {x_3})
\end{align*} through which \(A,B\) and \(C\) to be determined
to eliminate the second-order derivatives in \(\tilde{H}\). Thus it
will be a first-order differential operator equal to \(H\) on
\(\FSpace{L}{\phi}(\Space{H}{}/Z)\). To achieve this we need to take
\(A=\frac{\rmi}{4\pi m}\), \(B=-\frac{\rmi}{4\pi}\omega\), for
\(E=m\omega\). Note, that the value of \(E\) is uniquely defined and
consequently the corresponding vacuum vector \(\phi(y)=\rme^{-\pi \myhbar Ey^2}\)  is fixed. Furthermore, to obtain a geometric action
of \(\tilde{H}\) in the Schr\"odinger equation we need
pure imaginary coefficients of the first-order derivatives in
\(\tilde{H}\). Thus, this produces
\(C=-\frac{\rmi\myhbar \omega}{2} x_1\), with the final result:
\begin{align}
  \nonumber
  \tilde{H}&=\left(\frac{1}{2m}\left(\frac{\rmi\rmd\tilde{\sigma}_\myhbar^{X_1}}{\sqrt{2\pi}}\right)^2+\frac{m\omega^2}{2}\left(\frac{\rmi\rmd\tilde{\sigma}_\myhbar^{X_3}}{\sqrt{2\pi}}\right)^2\right) \\
  \nonumber
  &\quad  +(\frac{\rmi}{4\pi m}\partial_{1}+\frac{\omega}{4\pi}
  \partial_{3}-\frac{\rmi \myhbar\omega}{2} x_1 I)(-\rmi\linv {X_1}+ m\omega\linv
    {X_3})
    \\
  \label{eq:oscillation-rotations}
  &=\frac{ \rmi  \myhbar}{m} x_3 \partial_{1}- \rmi  \myhbar m \omega^2 x_1
\partial_{3}
    +  \left(\frac{1}{2} \myhbar \omega + \frac{\pi \myhbar^2}{m}(x_3^2-m^2\omega^2 x_1^2)\right) I\,.
\end{align}
Note, that operators~\eqref{eq:harmonic-hamilt}
and~\eqref{eq:oscillation-rotations} are not equal in general but have
the same restriction to the kernel of the auxiliary analytic
condition~\eqref{eq:cauchy-riemann}.  Thus, we aim to find a solution
to~\eqref{hami for harminc on Heisenberg group}, with \(\tilde{H}\)~\eqref{eq:oscillation-rotations} 
in place of \(H\)  on the
space of functions that satisfies the analyticity
condition~\eqref{eq:cauchy-riemann}. Since the operator
in~\eqref{eq:cauchy-riemann} (for \(E=m\omega\)) is intertwined with the Cauchy--Riemann
type operator by operator of multiplication by
\(\rme^{-\pi \myhbar m\omega x_1^2}\), the
generic solution of~\eqref{eq:cauchy-riemann} is
\begin{equation}
  \label{eq:generic-analytic-Heisenberg}
  f(x_1,x_3)=\rme^{-\pi\myhbar m\omega x_1^2} f_{1}\left(x_3- \rmi m\omega x_1\right)\,,
\end{equation}
where \(f_1(z)\) is an analytic function of one complex variable. Then, we use the method of characteristics  to solve the first-order PDE~\eqref{hami for
  harminc on Heisenberg group} with
\(\tilde{H}\)~\eqref{eq:oscillation-rotations} for the function \(f(x_1,x_3)\) \eqref{eq:generic-analytic-Heisenberg} to  obtain the solution:
\begin{equation}
  \label{final solution of harmonic ocs for Heisen}
  f(t;x_1,x_3)=\exp\left(\frac{-\rmi \omega}{2} t+ \pi{\rmi} \myhbar x_1x_3-\frac{\pi \myhbar}{2m\omega}(m^2\omega^2 x_1^2+x_3^2)\right) f_{2}\left(\rme^{ -\rmi \omega t}(x_3- \rmi m\omega x_1)\right)
\end{equation}
where for \(t=0\) the solution formula \eqref{final solution of harmonic ocs for Heisen} is a particular case of \eqref{eq:generic-analytic-Heisenberg} with \(f_1(z)=\rme^{-\frac{\pi \myhbar}{2m\omega}z^2}f_2(z)\) so it satisfies the analyticity condition \eqref{eq:cauchy-riemann}. Thus, formula~\eqref{final
  solution of harmonic ocs for Heisen} also satisfies~\eqref{hami for
  harminc on Heisenberg group} with the proper harmonic oscillator
Hamiltonian~\eqref{eq:harmonic-hamilt}. It is well known that this solution geometrically corresponds to a uniform rotation of the
phase space.


\subsection{Harmonic oscillator from the group $\Space{G}{}$}
\label{sec:harm-oscill-from-shear}

Here we obtain an exact solution of the harmonic oscillator in the
space \(\FSpace{L}{\phi}(\Space{G}{}/Z)\). It is fulfilled by the reduction of the order of the corresponding differential operator in a
manner illustrated on the Heisenberg group in
Section~\ref{sec:harm-oscill-heisenberg}. A new feature of this case is
that we need to use both operators~\eqref{eq:analyticity-anh}
and~\eqref{eq:general-annih} simultaneously. 

The Weyl quantisation of the harmonic oscillator Hamiltonian
\(\frac{1}{2}(p^2/m+m \omega^2 q^2)\) is given by:
\begin{align}\nonumber
 H&= \left( \frac{1}{2m}\left(\frac{\rmi\rmd\uir[2]{X_1}{\myhbar_4}}{\sqrt{2\pi}}\right)^2+\frac{m\omega^2}{2}\left(\frac{\rmi\rmd\uir[2]{X_3}{\myhbar_4}}{\sqrt{2\pi}}\right)^2 \right) \\
  \label{eq:hamiltonian-harm-an}
  &=-\frac{1}{4\pi m}\partial_{11}^{2} - \frac{1}{4\pi m} x_2^{2}\partial_{33}^{2}-\frac{m\omega^2}{4\pi} \partial_{33}^{2}-\frac{1}{2\pi m} x_2 \partial_{13}^{2} \\
  \nonumber\quad 
     &\quad +\frac{ \rmi \myhbar_4}{m} x_3\partial_{1}+\frac{ \rmi \myhbar_4}{m}x_2  x_3 \partial_{3}-\frac{1}{2 m} (- \rmi \myhbar_4x_2-2
       \pi\myhbar_4^{2} x_3^{2} ) I \,.
\end{align}
 Although, the Hamiltonian seems a bit alienated in comparison with the Heisenberg group case, cf.~\eqref{derived re reducibl re anharmo group}, we can still
adjust it by using conditions~\eqref{eq:analyticity-anh}
and~\eqref{eq:general-annih} as follows:
\begin{equation}
\label{adjusted eq}
  H_1=H+(A\partial_{1}+B\partial_{2}+C\partial_{3}+KI)\oper{C}+F\oper{S}\,.
\end{equation}
To eliminate all second order derivatives one has to take
\(A=\frac{\rmi}{4\pi m}\), \(B=0\), \(C=\frac{\rmi}{2\pi m}(\frac{-\rmi}{2} E+x_2)\)
and \(F=-\frac{1}{4\pi m}(\rmi x_2+ E)^2+\frac{m\omega^2}{4\pi}\). A
significant difference from the Heisenberg group case is that there is
no particular restrictions for the parameter \(E\). This allows to use
functions \(\rme^{-\pi \myhbar_4 E y^2}\), with any \(E>0\) as fiducial
vectors. Such functions are known as squeezed (or
two-photon~\cite{Yuen76}) 
 states and play an important role in quantum
theory, see original papers~\cites{WodkiewiczEberly85, Wodkiewicz87,
 Walls83, Stoler70, deGosson13a} and
textbooks~\citelist{\cite{Walls08}*{\S\,2.4--2.5}
  \cite{Schleich01a}*{\S\,4.3} \cite{Gazeau09a}*{Ch.~10}
 \cite{deGosson11a}*{\S\,11.3}}.

To make the action of the first order operator geometric we make
coefficients in front of the first-order derivatives \(\partial_{1}\)
and \(\partial_{3}\) imaginary. For this we put
\(K= \frac{\rmi\myhbar_4}{2m} x_1 (-E+2\rmi x_2)\). The final result is:
\begin{align}
   \label{eq:harmonic-1st-order}
 H_1&={H+(\frac{1}{4\pi m}(x_2+\rmi E)^2+\frac{m\omega^2}{4\pi})\oper{S}}\quad &\\
  \nonumber
  &\quad +\left(\frac{\rmi}{4\pi m}\partial_{1}+\frac{\rmi}{2\pi m}(\frac{-\rmi}{2}
    E+x_2)\partial_{3}+ \frac{\rmi \myhbar_4}{2m} x_1 (-E+2\rmi x_2)\right)\left(-\rmi
  \linv {X_1}+ E\linv {X_3}\right)\\
  \nonumber
     &= \frac{ \rmi \myhbar_4}{m}\left((x_3 + x_1 x_2) \partial_{1}
       -\left((\rmi x_2+ E)^2-m^2\omega^2\right)\partial_{2}
       -( E^{2} x_1-x_2 x_3 )  \partial_{3}  \right)\\
  \nonumber
  &\quad  -\frac{\myhbar_4}{2m}\left(-8 \rmi  \pi \myh_2  E x_2 -{ \rmi}  x_2 
    +4  \pi \myh_2  x_2^{2}
-2 \pi\myhbar_4 x_3^{2}\right.\\
  \nonumber
  &\qquad \left.   
    +4  \pi \myh_2 m^2\omega^2-  E -4  \pi \myh_2 E^{2} 
    -{4 \rmi}  \pi E  \myhbar_4x_1^{2} x_2 +2  \pi\myhbar_4 E^{2} x_1^{2} \right) I\,.
\end{align}
Thus,  the Schr\"odinger equation  for the harmonic oscillator 
\begin{equation}
  \label{schrodinger eq for an harm Hamilt}
  \rmi \myhbar_4\partial_{t}f(t;x_1,x_2,x_3)-H f(t;x_1,x_2,x_3)=0,
\end{equation}
is equivalent to the  first order linear PDE 
\begin{equation}
  \label{schrodinger eq for an harm Hamilt modified}
  \rmi\myhbar_4 \partial_{t}f(t;x_1,x_2,x_3)-H_1 f(t;x_1,x_2,x_3)=0,
\end{equation}
for \(f\) in the image space
\(\FSpace{L}{\phi}(\Space{G}{}/Z)\). It can be solved by the
following steps:
\begin{enumerate}
\item We start by \emph{peeling} the analytic operator~\eqref{eq:analyticity-anh} into a Cauchy--Riemann type operator. Indeed, the operator of multiplication
  \begin{displaymath}
   \oper{G}: f(x_1,x_2,x_3) \mapsto  \rme^{-\pi \myhbar_4E x_1^2} f(x_1,x_2,x_3)
 \end{displaymath}
 intertwines the analytic operator~\eqref{eq:analyticity-anh}---the
 Cauchy--Riemann type operator:
 \begin{displaymath}
   \oper{G}\circ \oper{C} = (E \partial_3-\rmi \partial_1) \circ \oper{G}. 
 \end{displaymath}
 Thus, a generic solution of the analytic
 condition \(\oper{C} f(x_1,x_2,x_3)=0\) has the form:
 \begin{equation}
   \label{eq:generic-soln-analytic-condition}
   f(x_1,x_2,x_3)= \rme^{-\pi \myhbar_4E x_1^2}  f_1(x_3-\rmi Ex_1,\rmi x_2+E),
 \end{equation}
 where \(f_1\) is an arbitrary analytic function of two complex variables.
\item Subsequently, the first order PDE~\eqref{schrodinger eq for an
    harm Hamilt modified} for the function
  \(f(x_1,x_2,x_3)\)~\eqref{eq:generic-soln-analytic-condition}
  has, by the standard method of characteristics, the following general solution:
  \begin{equation} 
  \label{solution to the shrodinger equation for shear case}
  \begin{split}
    f(t;x_1,x_2,x_3)&=\,\frac{\sqrt{E+m\omega}}{\sqrt{\rmi
        x_2+
        E+m\omega}}\\
    &\quad \times \exp\left(\frac{-\rmi \omega}{2} t-\pi{\myhbar_4} E
      x_1^2- 2\pi\rmi \myh_2 x_2 - \pi{\myhbar_4} \frac{(x_3-\rmi
        Ex_1)^2}{\rmi x_2+ E+m\omega}\right)
    \\
    &\quad \times f_2\left(\rme^{-\rmi \omega t}\frac{x_3-\rmi
        Ex_1}{\rmi x_2+ E+m\omega}, \rme^{-2\rmi \omega t}\frac{
        m\omega-(\rmi x_2+ E)}{m\omega+(\rmi x_2+E)}\right).
  \end{split}
  \end{equation}
   This is a particular case of formula ~\eqref{eq:generic-soln-analytic-condition} for \(t=0\)  and \[f_1(y_1,y_2)=\frac{\sqrt{E+m\omega}\exp\left(2\pi\rmi\myh_2( E-y_2)-\frac{\pi\myhbar_4y_1^2}{y_2+m\omega}\right)}{\sqrt{y_2+m\omega}}f_2\left(\frac{y_1}{y_2+m\omega},\frac{
        m\omega-y_2}{m\omega+y_2 }\right)\] here\, \(y_1=x_3-\rmi E x_1\) and\, \(y_2=\rmi x_2+E\) and \(f_2\) is an arbitrary analytic function of two complex variables. Hence, \eqref{solution to the shrodinger equation for shear case} satisfies the analyticity condition~\eqref{eq:analyticity-anh}.
  It also satisfies~\eqref{schrodinger eq for an harm Hamilt  modified} with the
  reduced Hamiltonian~\eqref{eq:harmonic-1st-order},
yet, it is not a
  solution to~\eqref{schrodinger eq for an harm Hamilt} with the
  original Hamiltonian~\eqref{eq:hamiltonian-harm-an}.
\item In the final step we request that~\eqref{solution to the shrodinger equation for
    shear case} satisfy the structural
  condition~\eqref{eq:general-annih}. This results in   a heat-like
  equation in terms of \(f_2\):
  \begin{equation}
    \label{eq:heat-like-f1}
    \partial_{u}f_2(z,u)=-\frac{1}{8\pi\myhbar_4  m\omega} \partial^{2}_{zz}f_2(z,u)
  \end{equation} where
  \begin{displaymath}
    z=\frac{x_3-\rmi Ex_1}{\rmi x_2+E+m\omega}\,,\quad  u=\frac{ m\omega-(\rmi x_2+ E)}{m\omega+(\rmi x_2+E)}\,.
  \end{displaymath}
  The well-known fundamental solution of~\eqref{eq:heat-like-f1} is:
  \begin{equation}
    \label{sol of heat-equation}
    f_2(z,u)=\left(\frac{2m\omega\myhbar_4}{u} \right)^{1/2}\int_{\Space{R}{}}g(\xi)\,\rme^{2\pi\myhbar_4 m\omega\frac{(z-\xi)^2}{u}}\, \rmd \xi,
  \end{equation}
  where \( g\) is the initial condition \(g(z)=f_2(z,0)\) and we use analytic extension from the real variable \(u\) to some
  neighbourhood of the origin in the complex plane---see discussion
  in \S\,\ref{sec:geom-phys-mean}. Thus, with \(f_2\) as in~\eqref{sol of heat-equation},
  formula~\eqref{solution to the shrodinger equation for shear case}
  yields a generic solution to~\eqref{schrodinger eq for an harm
    Hamilt} with Hamiltonian~\eqref{eq:hamiltonian-harm-an}.
\end{enumerate}
\begin{rem}
 Note that the above initial value of \(f_2\), related to \(u=0\), corresponds to \(x_2=0\) and \(E=m\omega\). Thus, it indicates that this general solution  is obtained from the unsqueezed states of Fock--Segal--Bargmann space, that is, the case related to the Heisenberg group. 
\end{rem}

\subsection{Geometrical, analytic and physical meanings of new
  solution}
\label{sec:geom-phys-mean}

Analysing the new solution~\eqref{solution to the shrodinger equation for shear case}
we immediately note that it is converted by the substitution \(x_2=0\)
(no shear) and \(E=m\omega\) (predefined non-squeezing for the Heisenberg
group) into the solution \eqref{final solution of harmonic ocs for Heisen}.  Thus, it is interesting to analyse the meaning of
formula~\eqref{solution to the shrodinger equation for shear case} for other
values. This can be deconstructed as follows.

\begin{figure}[htbp]
  \centering
\includegraphics[scale=.75]{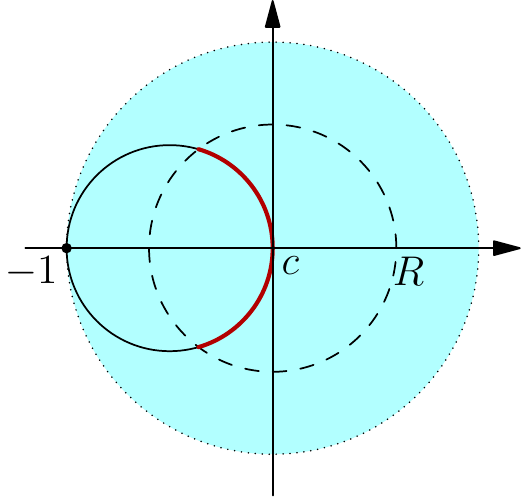}  
\includegraphics[scale=.75]{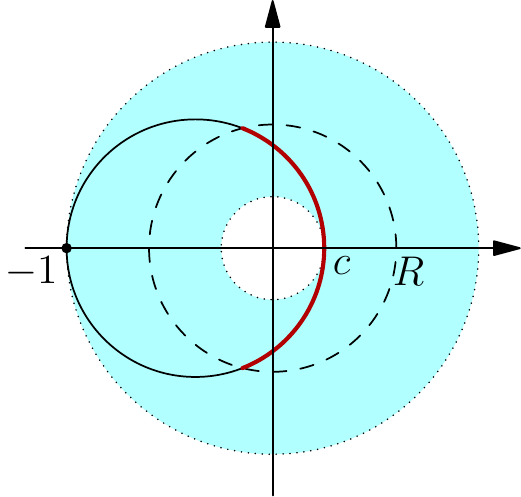}  
\includegraphics[scale=.75]{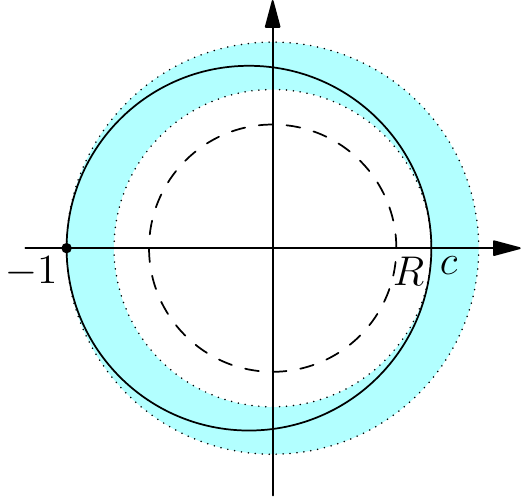}  
\caption[Shear parameter and analytic continuation]{Shear parameter
  and analytic continuation. The solid circle is the image of the line
  \(\rmi x_2+E\) under the Cayley
  type-transformation~\eqref{eq:Cayley-type}. The shadowed region with
  dotted boundary (the annulus with radii \(c\) and \(1\)) is obtained
  from the solid circle under rotation around the origin. The dashed
  circle of the radius \(R\) bounds the domain of the analytic
  continuation of the solution~\eqref{sol of
    heat-equation}.\\
  The left picture corresponds to \(E=m\omega\) (thus \(c=0\))---there
  always exists a part of the shaded region inside the circle of a
  radius \(R\) (even for \(R=0\)).\\
  The middle picture represents a case of some \(E\) within the
  bound~\eqref{eq:range-of-squeeze}---there is an arc (drawn by a
  thick pen) inside of the dashed circle. The arc corresponds to
  values of \(x_2\) such that the solution~\eqref{solution to the
    shrodinger equation for shear
    case} is meaningful.\\
  The right picture illustrates the shear parameter \(E\), which is
  outside of the range~\eqref{eq:range-of-squeeze}. For such a state,
  which is squeezed too much, no values of \(x_2\) allow to use the
  region of the analytic continuation within the dashed circle.}
  \label{fig:shear-param-analyt}
\end{figure}

The first factor, shared by any solution, is responsible for
\begin{enumerate}
\item  adding the value \(\frac{1}{2}\myhbar_4\omega\) to every integer
  multiple of \(\myhbar_4\omega\) eigenvalue;
\item peeling the second factor to analytic function;
\item proper \(\FSpace{L}{2}\)-normalisation. 
\end{enumerate}
The first variable in the function \(f_2\) of the second factor
produces integer multiples of \(\myhbar_4\omega\)  in eigenvalues. The
rotation dynamics of the second variable alliterate the shear
parameter as follows.  Points \(\rmi x_2+E\)  form a vertical line on the
complex plane. The Cayley-type transformation
\begin{equation}
  \label{eq:Cayley-type}
  \rmi x_2+E \ \mapsto \ \frac{m\omega-(\rmi x_2+E)}{m\omega+(\rmi x_2+ E)}
\end{equation}
maps the vertical line \(\rmi x_2+ E\) into
the circle with the centre \(\frac{-E}{m\omega+E}\) and radius \(\frac{m\omega}{m\omega+ E}\) (therefore passing \(-1\)). Rotations of  a point of this circle around the origin creates circles centred at the origin
and a radius between \(c=\modulus{\frac{m\omega-
  E}{m\omega+ E}}\) and \(1\), see Fig.~\ref{fig:shear-param-analyt}.

Let a function \(f_2\) from~\eqref{sol of heat-equation} have an analytic
extension from the real values of \(u\) into a (possibly punctured)
neighbourhood of the origin of a radius \(R\) in the complex
plane. An example of functions admitting such an extension are the eigenfunctions of the harmonic
oscillator~\eqref{eq:eigenfunctions-shear-group} considered in the
next section. In order for the solution~\eqref{solution to the shrodinger
  equation for shear case} to be well-defined for all values of \(t\)
one needs to satisfy the inequality
\begin{equation}
  \label{eq:range-of-oscillations}
  \modulus{\frac{m\omega-(\rmi x_2+
  E)}{m\omega+(\rmi x_2+ E)}}< R\,.
\end{equation}
It implies the allowed range of \(E\) around the special value
\(m\omega\):
\begin{equation}
  \label{eq:range-of-squeeze}
  \frac{1-R}{1+R} m \omega < E < \frac{1+R}{1-R}m\omega.
\end{equation}
For every such \(E\), the respective allowed range of \(x_2\) around
\(0\) can be similarly deduced from the required
inequality~\eqref{eq:range-of-oscillations}, see the arc drawn by
thick pen on Fig.~\ref{fig:shear-param-analyt}.

The existence of bounds~\eqref{eq:range-of-squeeze} for possible
squeezing parameter \(E\) shall be expected from the physical
consideration. The integral formula~\eqref{sol of heat-equation}
produces for a real \(\rmi u\) a solution of the irreversible
heat-diffusion equation for the time-like parameter \(u\).  However,
its analytic extension into the complex plane will include also
solutions of time-reversible Schr\"odinger equation for purely
imaginary \(\rmi u\).  Since the rotation of the second variable
in~\eqref{solution to the shrodinger equation for shear case} requires
all complex values of \(u\) with fixed \(\modulus{u}\), only a
sufficiently small neighbourhood (depending on the ``niceness'' of an
initial value \(f_2(z,0)\)) is allowed. Also note, that a rotation of
a squeezed state in the phase space breaks the minimal uncertainty
condition at certain times, however the state periodically
``re-focus'' back to the initial minimal uncertainty
shape~\cite{Wodkiewicz87}.

If a solution \(f_2(z,u)\), \(u\in\Space{R}{}\) of~\eqref{sol of
  heat-equation} does not permit an analytic expansion into a
neighbourhood of the origin, then two analytic extensions  \(f^{\pm
}_2(z,u)\), for \(\Im u>0\) and  \(\Im u<0\) respectively, shall
exist. Then, the dynamics
in~\eqref{solution to the shrodinger equation for shear case} will experience two
distinct jumps for all values of \(t\) such that
\begin{equation}
  \label{eq:jump-condition}
  \Re\left(\rme^{-2 \rmi \omega t}\frac{ m\omega-(\rmi x_2+ E)}{m\omega+(\rmi x_2+ E)}\right)=0\,.
\end{equation}
An analysis of this case and its physical interpretation is beyond the
scope of the present paper.
\section{Ladder operators on the group $\Space{G}{}$}
For determining a complete set of eigenvectors of Hamiltonian ~\eqref{eq:hamiltonian-harm-an} we consider ladder operators.
Let
\begin{displaymath}
a^+=\sqrt{\frac{m\omega}{ \pi\myhbar_4}}\left(\frac{\rmi}{2m\omega} X_1- \frac{1}{2} X_3\right) \quad \text{and}\quad  a^ -=\sqrt{\frac{m\omega}{ \pi\myhbar_4}}\left(\frac{\rmi}{2m\omega} X_1+ \frac{1}{2} X_3\right).
\end{displaymath}
Using the derived representation formulae~\eqref{derived re reducibl
  re anharmo group} we have the following dimensionless operators: 
\begin{align*}
L^{+}:=\rmd\uir[2]{a^+}{\myh_4}&=\sqrt{\frac{m\omega}{\pi\myhbar_4}}\left(\frac{\rmi}{2m\omega} \rmd\uir[2]{X_1}{\myhbar_4} - \frac{1}{2}\rmd\uir[2]{X_3}{\myhbar_4}\right) \\&
= \frac{\rmi}{2\sqrt{\pi m\omega\myhbar_4}}(- \partial_{1}-( x_2+\rmi m\omega)\partial_{3}+ 2\pi\rmi \myhbar_4  x_3\,  I);\\
L^{-}:=\rmd\uir[2]{a^ -}{\myhbar_4}&=\sqrt{\frac{m\omega}{\pi\myhbar_4}}\left(\frac{\rmi}{2m\omega} \rmd\uir[2]{X_1}{\myhbar_4} + \frac{1}{2}\rmd\uir[2]{X_3}{\myhbar_4}\right)\\
&= \frac{\rmi}{2\sqrt{\pi m\omega\myhbar_4}}(-\partial_{1}-( x_2-\rmi m\omega)\partial_{3}+2\pi \rmi \myhbar_4  x_3 \, I)\,.
\end{align*}
One can immediately verify the commutator
\begin{equation}
\label{commuation of ladders}
[L^-,L^+]= I\,.
\end{equation}
 The Hamiltonian~\eqref{eq:hamiltonian-harm-an} is expressed in terms of the ladder operators:
 \begin{equation}
 \label{Hami in terms of ladder}
  H=\myhbar_4\omega (L^+ L^-+\frac{1}{2}\,I)\,.
  \end{equation}
Thus the following commutators hold:
  \begin{equation*}
\label{hamiltonian and ladder}
[H,L^+]= \myhbar_4\omega L^+, \quad  [H,L^-]=- \myhbar_4\omega L^-.
\end{equation*}
Moreover, creation and annihilation operators are adjoint of each other:
\begin{equation}
\label{adjoint ladder}
(L^-)^*=L^{+}
\end{equation}
where `*' indicates the adjoint of an operator in terms of the inner product defined by~\eqref{inner product of the image}.
Then, from~\eqref{Hami in terms of ladder} and~\eqref{adjoint ladder} we  see that \(H^*=H\).

Now, for the function (see~\eqref{wavelet of mother wavelet shear group} for \(q=m\omega\))
\begin{equation}
\label{the vacuum state}
\Phi_0(x_1,x_2,x_3)=\frac{\sqrt{2}(m\omega E)^{1/4}}{\sqrt{ \rmi x_2+ E+m\omega}}
\exp \left(-\pi{\myhbar_4} Ex_1^2- 2\pi\rmi \myh_2x_2-\pi{\myhbar_4}\frac{(x_3 -\rmi E x_1)^2}{ \rmi x_2+ E+m\omega}\right),
\end{equation} it can be easily
checked that it is a vacuum vector in
\(\FSpace{L}{\phi}(\Space{G}{}/Z)\):
\begin{equation*}
 L^-\Phi_0=0.
\end{equation*} It is normalised
(\(
\norm[x_2]{\Phi_0}=1
\))
and for the higher order normalised states, we put
\begin{displaymath}
\Phi_j=\frac{1}{\sqrt{j!}}(L^+)^j\Phi_0,
\quad \quad j=1,2,\hdots.
\end{displaymath} Orthogonality of \(\Phi_j\) follows from the fact that \(H\) is self-adjoint.
 Then, for
\begin{displaymath}
z=\frac{x_3-\rmi E x_1}{\rmi x_2+ E+m\omega},\quad  u=\frac{ m\omega-(\rmi x_2+ E)}{m\omega+(\rmi x_2+E)}
\end{displaymath}
we have
\begin{equation}
  \label{eq:eigenfunctions-shear-group}
\Phi_j(z,u)=\frac{1}{\sqrt{2^j j!}}(- u)^{j/2} H_j\left(\sqrt{\frac{-2\pi\myhbar_4m\omega}{u}}\;z\right)\Phi_0
\end{equation}
where
\begin{equation}
  \label{eq:hermite}
  H_{j}(y)=\sum_{k=0}^{\lfloor \frac{j}{2}\rfloor}\frac{(-1)^k j!}{k!(j-2k)!}(2y)^{j-2k}
\end{equation}
are the \emph{Hermite polynomials}
~\citelist{\cite{Folland89}*{\S\,1.7} \cite{Andrews98a}*{\S\,5.2}}
 and \( \lfloor\cdot\rfloor\) is the floor function (i.e. for any real \(x\), \( \lfloor x\rfloor\) is the greatest integer \(n\) such that \(n\leq x\).) And,
 \begin{align}
 \label{vacuum in terms of z and u}
 \nonumber
 \Phi_0(z,u)&=\left( \frac{E}{m\omega}\right)^{1/4}\rme^{2\pi \myh_2 E}\sqrt{1-u}\exp\left(\frac{\pi\myhbar_4E m^2\omega^2}{(1-u)^2}\big(z+(\frac{E}{m\omega}(-1+ u)+ u)\overline{z}\big)^{2}\right)\\
 &\qquad\times\exp\left(\frac{-2\pi m\omega}{1-u}\big(\myh_2{(1+u)}+\myhbar_4 {z^2}\big) \right).
 \end{align}
 
 It is easy to show (by induction in which we use~\eqref{commuation of ladders}) that
\begin{align}
\label{1}
L^- \Phi_j= [L^-,\frac{1}{\sqrt{j!}}(L^+)^j]\Phi_0=\sqrt{j} \Phi_{j-1}.
\end{align}
Therefore,
\begin{equation*}
 L^+L^-\Phi_j= j \Phi_j.
\end{equation*} Hence,
 for the Hamiltonian \(H\) of the harmonic oscillator ~\eqref{eq:hamiltonian-harm-an} we have
\begin{equation*}
\textstyle H\Phi_j=\myhbar_4\omega(j+\frac{1}{2})\ \Phi_j.
\end{equation*}

Furthermore, it can be verified that both operators \(\oper{C}\) \eqref{eq:analyticity-anh} and \(\oper{S}\) \eqref{eq:general-annih} commute with the creation operator \(L^+\) and thus 
\begin{equation*}
\oper{C} \Phi_j=\oper{S}\Phi_j=0, \quad j=0,1,2,\hdots.
\end{equation*}

Note, that singularity of
eigenfunction~\eqref{eq:eigenfunctions-shear-group} at \(u=0\) is
removable due to a cancellation between the first power factor and the
Hermite polynomial given by~\eqref{eq:hermite}. Moreover, the
eigenfunction~\eqref{eq:eigenfunctions-shear-group} has an analytic
extension in \(u\) to the whole complex plane, thus does not have any
restriction on the squeezing  parameter \(E\) from the
inequality~\eqref{eq:range-of-oscillations}.

The eigenfunction
\(\Phi_j(z,u)\)~\eqref{eq:eigenfunctions-shear-group} at \(u=0\) 
reduces to the power \(z^j\) of the variable \(z\), as can be expected
from the connection to the FSB space and the Heisenberg
group. The appearance of the Hermite polynomials in
\(\Phi_j(z,u)\)~\eqref{eq:eigenfunctions-shear-group} may be a bit
confusing since the Hermite functions represent eigenvectors in the
Schr\"odinger representation over
\(\FSpace{L}{2}(\Space{R}{})\). However, if we  substitute the
dynamic~\eqref{solution to the shrodinger equation for shear case}: \begin{equation*}
z_t=\frac{x_3-\rmi Ex_1}{\rmi x_2+ E+m\omega}\rme^{ -\rmi\omega t},\qquad u_t=\frac{ m\omega-(\rmi x_2+ E)}{m\omega+(\rmi x_2+E)}\rme^{-2
          \rmi \omega t}
\end{equation*} into the
eigenfunction~\eqref{eq:eigenfunctions-shear-group}
 we get
\begin{align*}
\Phi_j(z_t,u_t)= \frac{1}{\sqrt{2^j j!}}(- u_t)^{j/2}\,H_j\bigg(\sqrt{\frac{-2\pi\myhbar_4m\omega}{u}}\;z\bigg)\Phi_0(z_t,u_t)\,.
\end{align*}
Thus, the argument of Hermite function ``stays still'' while the time
parameter is present only in the power factor and the vacuum~\eqref{vacuum in terms of z and u}. This is completely inline
with the FSB space situation.

\section{Discussion and conclusions}

We presented a method to obtain a geometric solution of Schr\"odinger
equation expressed through coordinate transformation. The method
relays on coherent state transform based on group
representations~\citelist{\cite{Perelomov86} \cite{AliAntGaz14a}*{Ch.~7}}.
 It is shown that properties of the
solution depends on a group and its representation used in the
transform. The comparison of the Heisenberg and shear \(\Space{G}{}\)
groups cases shows that a larger group \(\Space{G}{}\) creates the
image space with a bigger number of auxiliary conditions. These
conditions can be used to reduce the order of a partial differential
equation with bigger flexibility, leading to a richer set of geometric
solutions. We have seen that the fiducial vector for the Heisenberg
group is uniquely defined while for the larger group \(\Space{G}{}\) different minimal
uncertainty states still lead to a geometric solution. There are some
natural bounds~\eqref{eq:range-of-squeeze} of a
possible squeeze parameter, they are determined by the degree of
singularity of the solution of the
equation~\eqref{eq:heat-like-f1}. It would be interesting to find a
physical interpretation of jumps for values~\eqref{eq:jump-condition}
for states which do not have an analytic continuation into a
neighbourhood of the origin.  

The present work provides a further example of numerous
cases~\cites{Kisil09d,Kisil10c,Kisil98a,Zimmermann06,Guerrero18} when
the coherent state transform is meaningful and useful beyond the
traditional setup of square-integrable representations modulo a
subgroup \(H\)~\cite{AliAntGaz14a}*{Ch.~8}. Specifically, coherent
states parametrised by points of the homogeneous space \(\Space{G}{}/H\)
are not sufficient to accommodate the dynamics~\eqref{solution to the
  shrodinger equation for shear case}.

It would be interesting to continue the present research for other
groups.  One of the immediate candidates was suggested to us by an
anonymous referee: the Newton--Hook group~\cites{BacryLevy-Leblond68a,Streater67},
which is the Heisenberg group added with time translations generated
by the harmonic oscillator Hamiltonian. A more refined coherent state
transform can be achieved by the Schr\"odinger group \(\Space{S}{}\)
introduced in \S\,\ref{sec:shear-group-schr} because it is possibly
the largest natural group for describing coherent states for the
harmonic oscillator, see
\citelist{\cite{AldayaCossioGuerreroLopez-Ruiz11b}
  \cite{AldayaGuerrero01a} \cite{Folland89}*{Ch.~5}}. However, as was
mentioned at the end of \S\,\ref{sec:shear-group-schr}, the smaller
group \(\Space{G}{}\) has more representations than the larger
Schr\"odinger group. Thus advantages of each group for geometric
description of dynamics needs to be carefully investigated.

\section*{Acknowledgments}
\label{sec:acknowledgments}
The first-named author was supported by a scholarship from the Taif
University (Saudi Arabia).  Authors are grateful to Prof.~S.M.\;Sitnik
for helpful discussions of Gauss-type integral operators. The visit of
Prof.~S.M.\;Sitnik to Leeds was supported by the London Mathematical
Society grant (Scheme 4). Prof.\;M.\;Ruzhansky pointed out that the group
\(\Space{G}{}\) belongs to the family of Engel groups and
Prof.\;R.\;Cam\-po\-amor-Sturs\-berg kindly informed us about filiform algebras
considered in~\cite{Vergne70a}.  Authors are very grateful
to anonymous referees for many useful comments and suggestions.


\providecommand{\noopsort}[1]{} \providecommand{\printfirst}[2]{#1}
  \providecommand{\singleletter}[1]{#1} \providecommand{\switchargs}[2]{#2#1}
  \providecommand{\irm}{\textup{I}} \providecommand{\iirm}{\textup{II}}
  \providecommand{\vrm}{\textup{V}} \providecommand{\cprime}{'}
  \providecommand{\eprint}[2]{\texttt{#2}}
  \providecommand{\myeprint}[2]{\texttt{#2}}
  \providecommand{\arXiv}[1]{\myeprint{http://arXiv.org/abs/#1}{arXiv:#1}}
  \providecommand{\doi}[1]{\href{http://dx.doi.org/#1}{doi:
  #1}}\providecommand{\CPP}{\texttt{C++}}
  \providecommand{\NoWEB}{\texttt{noweb}}
  \providecommand{\MetaPost}{\texttt{Meta}\-\texttt{Post}}
  \providecommand{\GiNaC}{\textsf{GiNaC}}
  \providecommand{\pyGiNaC}{\textsf{pyGiNaC}}
  \providecommand{\Asymptote}{\texttt{Asymptote}}

\end{document}